\documentclass[pra, 10pt, twocolumn, aps, superscriptaddress, showpacs, groupaddress]{revtex4-2}

\usepackage{natbib}
\usepackage{bigints}
\usepackage{amssymb}
\usepackage{graphicx}
\usepackage{amsmath}
\usepackage{dcolumn}
\usepackage{bm}
\usepackage{multirow}
\usepackage{hyperref}
\usepackage{physics}
\usepackage{bm}
\usepackage{comment}
\usepackage{xcolor}
\usepackage[caption=false]{subfig}
\usepackage{float}
\usepackage{siunitx}
\usepackage{dsfont}
\usepackage{tikz}

\hypersetup{colorlinks=true, linkcolor=blue, linktoc=page, citecolor=blue, urlcolor=blue}

\begin{document}

\title{Landau damping and mode-selective instability of a vortex gas}

\author{H. Ter\c cas}
\email{hugo.tercas@tecnico.ulisboa.pt}
\affiliation{Instituto Superior de Engenharia de Lisboa, Instituto Politécnico de Lisboa, 1959-007 Lisboa, Portugal}
\affiliation{GoLP/IPFN, Instituto Superior T\'ecnico, Universidade de Lisboa, 1049-001 Lisboa, Portugal}

\author{J. L. Figueiredo}
\affiliation{GoLP/IPFN, Instituto Superior T\'ecnico, Universidade de Lisboa, 1049-001 Lisboa, Portugal}

\author{J. T. Mendon\c{c}a}
\email{titomend@tecnico.ulisboa.pt}
\affiliation{GoLP/IPFN, Instituto Superior T\'ecnico, Universidade de Lisboa, 1049-001 Lisboa, Portugal}

\date{\today}

\begin{abstract}

We show that a collisionless point-vortex gas behaves as a genuine Vlasov system. Starting from the non-canonical Hamiltonian structure of point-vortex dynamics, we derive a reduced Klimontovich equation and the corresponding Vlasov description directly on the physical plane, without introducing auxiliary momentum variables. The theory predicts a rich spectrum of collective phenomena, including vortex Rossby-like waves, global oscillations, resonant critical layers, and a dielectric response operator analogous to that of plasmas and self-gravitating systems. For smooth vortex distributions, differential rotation gives rise to a vortex analogue of Landau damping, whereby collective modes transfer energy and angular momentum to resonant vortex annuli in the absence of collisions or dissipation. Remarkably, differential rotation acts as a stringent mode-selection mechanism: for a Gaussian vortex cloud, only a single quadrupolar quasi-mode survives, while all higher multipoles are destroyed by phase mixing. We further predict a vortex counterpart of the beam--plasma instability, in which a thin vortex ring orbiting an Onsager patch selectively excites Kelvin modes with an azimuthal symmetry controlled by the ring population. These results establish kinetic theory as a natural framework for collective vortex dynamics and suggest new mechanisms for vortex transport, vortex avalanches, and rotational glitches in neutron-star superfluids.
\end{abstract}

\maketitle

\section{Introduction}

Quantized vortices are among the most central physical manifestations of superfluids and superconductors, and they play a central role in quantum turbulence and non-equilibrium transport in superfluid systems
\cite{sonin,fetter,pethick,stringari}. The study of turbulent vortex configurations has, in particular, become an important component of modern quantum hydrodynamics, connecting the physics of ultracold gases turbulence and strongly nonlinear
many-body dynamics \cite{wave,pd2006}.

At length scales much larger than the core size, vortices may be regarded as topological point defects interacting through long-range logarithmic potentials, an observation that leads to the concept of a {\it vortex gas}. In a seminal work, Onsager showed that a gas of point vortices can sustain negative-temperature equilibrium states, allowing large-scale coherent structures to emerge spontaneously from microscopic disorder \cite{onsager}, thereby establishing a connection between two-dimensional turbulence and equilibrium statistical mechanics that remains highly influential today \cite{eyink_2006, bouchet2012, eyink_2024}. The statistical mechanics of point-vortex systems has since evolved into a broad
research area spanning fluid mechanics, plasma physics, geophysical flows and superfluidity
\cite{joyce,montgomery,miller,robert,sommeria,newton}, and the discovery of vortex clustering and inverse energy cascades in
two-dimensional quantum turbulence has revived interest in Onsager's
original predictions, which have now found experimental and numerical
support in atomic condensates \cite{neely,reeves,simula,bradley}. Such developments have highlighted the need for a systematic kinetic
description of vortex ensembles, capable of connecting microscopic
dynamics with collective behavior, and any program of that kind
immediately encounters a fundamental difficulty. Unlike ordinary
particles, point vortices do not obey Newtonian dynamics: their
equations of motion are first order in time and possess a non-canonical Hamiltonian structure in which the Cartesian coordinates themselves form a conjugate pair \cite{kirchhoff,marsden}. The natural phase space of a vortex gas is consequently not the conventional particle phase space spanned by independent position and momentum coordinates, so that any kinetic theory built on an enlarged phase space containing auxiliary momentum variables necessarily introduces redundant degrees of freedom and obscures the underlying symplectic geometry of the problem.

The kinetic theory of systems with long-range interactions is
traditionally based on the Klimontovich--Vlasov formalism
\cite{klimontovich}, which has proven extraordinarily successful in plasma physics \cite{nicholson}, self-gravitating systems \cite{heyvaerts2010} and non-neutral fluids \cite{dean_1996, demery_2016}. For point vortices, the subject is by now a mature one. Dubin and O'Neil obtained a Lenard--Balescu equation for axisymmetric vortex flows \cite{dubinoneil1988}, while a Landau-type equation was derived independently by Chavanis \cite{chavanis2001,chavanis2008} and later extended to include collective screening \cite{chavanis2012}. The angle--action formulation of these theories was developed for general inhomogeneous long-range systems by Heyvaerts \cite{heyvaerts2010} and Chavanis \cite{chavanis2012aa}, and it is this apparatus that we shall use in what follows. The formal parallel between the kinetic theories of point vortices and of stellar systems, in which the angular velocity profile plays the role of the orbital frequency profile and the systematic drift of a test vortex that of dynamical friction, has been developed at length by the same authors \cite{chavanis2024}; on the Hamiltonian side, the non-canonical Poisson structure underlying two-dimensional ideal hydrodynamics is likewise classical \cite{marsden,morrison1998}. 

The existing body of work is, in essence, a theory of granularity in
the number of vortices $N_v$. Whether collective polarization is
retained at $\mathcal{O}(1/N_v)$ or three-body correlations drive the
evolution at $\mathcal{O}(1/N_v^2)$ \cite{chavanis2012}, what is
computed is the slow relaxation of the coarse-grained profile toward a
maximum-entropy state, an effect that vanishes together with the
discreteness producing it. Important physics nonetheless survives the limit $N_v\to\infty$, since the mean-field contributions remain, and with it the collective response of the vortex gas. In plasma physics, such response constitutes the whole of the linear theory, carrying Landau damping and beam-driven instability, with the collision operator added beneath it only to account for the slow drift. Showing that a collisionless vortex gas damps and destabilizes in a similar manner, i.e. via particle-wave interaction, is the purpose of the present work. These features are expected to play an important role in rotating neutron stars $-$ e.g. the Vela pulsar \cite{dodson_2003}$-$, where arrays of $N_v\sim 10^{17}$ vortices in the superfluid nuclear matter are produced due to fast rotation \cite{haskellsedrakian,Antonopoulou2022}. With a discreteness parameter of that size, the collisional time exceeds the characteristic age of the star by several orders of magnitude, and every observable feature of the rotational dynamics, from the sub-minute rise of a glitch to the inter-glitch recovery, unfolds deep inside the collisionless window.

The paper is organized as follows. Section~II reviews the non-canonical Hamiltonian structure of point-vortex dynamics, derives the corresponding microscopic Klimontovich equation and establishes the reduced collisionless Vlasov equation on which everything that follows rests. Section~III turns to collective excitations, discussing the volume modes and their Rossby-wave character together with the global modes of two representative equilibria. Section~IV develops the theory of vortical Landau damping and of the resonant critical layers
responsible for it, while Sec.~V analyzes the instability driven by a vortex ring orbiting an Onsager patch. Section~VI summarizes our conclusions and discusses the implications for vortex transport in neutron-star superfluidity.

\section{The Klimontovich equation}

We consider a gas of $N_v$ parallel vortex lines, all oriented along the
$z$ direction. Restricting the dynamics to the perpendicular plane, the
velocity induced by the $\alpha-$th vortex with circulation
$\Gamma_\alpha=\oint{\bf u}_\alpha\cdot d{\bm\ell}=h/(2M)$ (with $h$ and $M$ respectively denoting Planck's constant and the
mass of the superfluid particles) located at position
${\bf r}_\alpha=(r^1_\alpha, r^2_\alpha)\equiv (x_\alpha,y_\alpha)$, is given by
\begin{equation}
{\bf u}_\alpha({\bf r})=\frac{\Gamma_\alpha}{2\pi} {\bf e}_z \times \frac{{\bf r}-{\bf r}_\alpha} {|{\bf r}-{\bf r}_\alpha|^2}.
\label{2.1}
\end{equation}
The total velocity field follows from the superposition of all vortex contributions and the corresponding interaction energy may be written in Hamiltonian form as \cite{newton,marsden}
\begin{equation}
H = -\frac{1}{4\pi} \sum_{\alpha\neq \beta}^{N_v}\Gamma_\alpha\Gamma_\beta\ln |{\bf r}_\alpha-{\bf r}_\beta|, 
\label{2.2}
\end{equation}
where the logarithmic interaction is a direct consequence of the Green
function of the two-dimensional Laplace operator satisfying
$\nabla^2\ln r=2\pi\delta({\bf r})$. The dynamics generated by
Eq.~\eqref{2.2} differs fundamentally from that of ordinary particle
systems \cite{kirchhoff,newton}, since point vortices possess no
independent momentum variables and their Cartesian coordinates instead
form a conjugate pair, satisfying the non-canonical Poisson bracket
\begin{equation}
\left\{r^i_\alpha,\,r^j_\beta\right\}
=\frac{J^{ij}}{\Gamma_\alpha}\,\delta_{\alpha\beta},
\label{2.5}
\end{equation}
where $J^{ij}={\rm offdiag}(1,-1)$ is the symplectic matrix such that $p_i^\alpha=\sum_j J_{ij}r_j^\alpha$ is the conjugate momentum. The equations of motion then follow from Hamilton's equations, $\dot{r}_\alpha^i =\{r_\alpha^i,H\}$, which, in more explicit terms, read
\begin{equation}
\dot{x}_\alpha=
\frac{1}{\Gamma_\alpha}
\frac{\partial H}{\partial y_\alpha},
\qquad
\dot{y}_\alpha=
-\frac{1}{\Gamma_\alpha}
\frac{\partial H}{\partial x_\alpha}.
\label{2.7}
\end{equation}
These relations indicate that the physical plane occupied by the
vortices already constitutes the natural phase space of the system, so that no auxiliary momentum coordinate is required. It is this property that distinguishes vortex gases from conventional kinetic systems, and it will play a central role in what follows.

Following Klimontovich's approach, we first introduce the microscopic
phase-space density
\begin{equation}
\mathcal N({\bf r},{\bf p},t)
=
\sum_{\alpha=1}^{N_v}
\delta({\bf r}-{\bf r}_\alpha)
\delta({\bf p}-{\bf p}_\alpha).
\end{equation}
The symplectic relation defined in Eq. \eqref{2.5} can be recast as ${\bf p}=\mathbb J\cdot{\bf r}$, such that integration over the constrained momentum yields
\begin{equation}
N({\bf r},t)=\int d{\bf p}\,
\mathcal N({\bf r},{\bf p},t)
 = \sum_{\alpha=1}^{N_v} \delta({\bf r}-{\bf r}_\alpha),
\label{2.8}
\end{equation}
which denotes the density field on the physical plane, 
reflects the reduced dimensionality of the vortex phase space. Taking the time derivative of Eq.~\eqref{2.8} and using the equations of motion \eqref{2.7}, we obtain
\begin{equation}
\frac{\partial N}{\partial t} = - \sum_{\alpha=1}^{N_v} \dot{\bf r}_\alpha \cdot \bm \nabla \delta \left( {\bf r}-{\bf r}_\alpha \right),
\label{2.9}
\end{equation}
which can be written as a continuity equation 
\begin{equation}
\frac{\partial N }{\partial t} + {\bm \nabla} \cdot \left( N {\bf u} \right) = 0,
\label{2.10}
\end{equation}
where ${\bf u}({\bf r},t)=\bm\nabla\psi({\bf r},t)\times \hat{\bf z}$
denotes the velocity field generated by all vortices, conveniently
expressed in terms of the microscopic stream function
\begin{equation}
\psi({\bf r},t) = -\frac{1}{2\pi} \int \Gamma({\bf r}') N ({\bf r}',t) \ln \vert {\bf r}-{\bf r}' \vert d^2r'.
\label{2.11}
\end{equation}
Equation \eqref{2.10} has been employed in the point-vortex and two-dimensional turbulence literature as the microscopic vorticity distribution \cite{newton}; here, however, we reinterpret it as the fundamental microscopic state variable associated with the reduced
phase space of a non-canonical vortex gas. For identical vortices of circulation $\Gamma$, which is the situation of interest throughout, the stream function reduces to
\begin{equation}
\psi({\bf r},t) = -\frac{\Gamma}{2\pi} \int N ({\bf r}',t) \ln \vert {\bf r}-{\bf r}' \vert d^2r',
\label{2.12}
\end{equation}
and, using the logarithmic Green's function, may be recast in the form
of a vortex Poisson equation,
\begin{equation}
\nabla^2\psi = -\Gamma N .
\label{2.13}
\end{equation}
Introducing the Poisson bracket
\begin{equation}
\{A,B\}
=
\frac{\partial A}{\partial x}
\frac{\partial B}{\partial y}
-
\frac{\partial A}{\partial y}
\frac{\partial B}{\partial x},
\label{2.16}
\end{equation}
we then arrive at the compact form of Eq.~\eqref{2.10},
\begin{equation}
\frac{\partial N}{\partial t} + \{N ,\psi\}=0.
\label{2.17}
\end{equation}
Equations \eqref{2.13} and \eqref{2.17} constitute the exact kinetic
description of a point-vortex gas. Being fully equivalent to the
microscopic equations of motion, they provide the natural starting
point for the construction of a kinetic theory, and, unlike
conventional Klimontovich formulations, they contain no independent
momentum coordinate, the resulting dynamics being governed entirely by
the non-canonical Hamiltonian structure of the vortex gas.

We now consider the coarse-grained vortex density,
\begin{equation}
f({\bf r},t)=\langle N({\bf r},t)\rangle,
\end{equation}
where $\langle \cdot \rangle$ denotes the ensemble average. Decomposing the microscopic quantities into mean and fluctuating components, $N=f+\delta N$ and $\psi=\Psi+\delta\psi$, and averaging the Klimontovich equation, one obtains
\begin{equation}
\frac{\partial f}{\partial t} + \left\{f,\Psi \right\}= -\langle \{\delta N,\delta\psi\} \rangle \equiv \mathcal{C}[f],
\label{eq:vlasov1}
\end{equation}
with $\mathcal{C}[f]\sim \mathcal{O}(\delta N^2)$ representing the effect of correlations between microscopic density fluctuations. It is the vortex analog of the collision integral appearing in the Klimontovich description of plasmas and self-gravitating systems \cite{klimontovich,nicholson,binney}. Since the fluctuating stream function is related to the fluctuating vortex density through
\begin{equation}
\delta\psi({\bf r},t)=-\frac{\Gamma}{2\pi}\int\delta N({\bf r}',t)\ln|{\bf r}-{\bf r}'|\, d^2r',
\label{3.8}
\end{equation}
the collision operator reads explicitly
\begin{equation}
\mathcal{C}[f]=\frac{1}{2\pi}\left\langle\left\{\delta N({\bf r}),\int \delta N({\bf r}') \ln|{\bf r}-{\bf r}'| \,d^2r' \right\} \right\rangle .
\label{3.9}
\end{equation}
This operator encapsulates the effect of discrete two-body and higher-order correlations between vortices, and in a systematic kinetic treatment it is evaluated by truncating the corresponding BBGKY-like
hierarchy through an asymptotic expansion in the discreteness parameter
$1/N_v$. At order $\mathcal{O}(1/N_v)$, retaining collective
polarization yields the vortex analogue of the Lenard--Balescu
equation, whereas neglecting such screening effects leads to a
Landau-type collision operator
\cite{dubinoneil1988,chavanis2001,chavanis2008}. In highly symmetric configurations, where the $\mathcal{O}(1/N_v)$ contribution vanishes identically, the evolution is driven instead by three-body correlations at order $\mathcal{O}(1/N_v^2)$ \cite{chavanis2012}. Those
works address the {\it secular} evolution of the vortex gas, i.e. the slow collisional relaxation of the coarse-grained profile toward its maximum-entropy state, whereas our aim here is physically distinct, being the fast collective dynamics, the resonances and the instabilities that occur well before collisional relaxation sets in.

To formalize this separation of timescales, we consider the mean-field (Vlasov) limit, taking $N_v,A\rightarrow\infty$ while keeping both the vortex density $N_v/A$ and the total circulation $\Gamma_{\rm tot}=N_v\Gamma$ fixed, so the circulation carried by each vortex therefore scales as $\Gamma=\Gamma_{\rm tot}/N_v$. For a finite but large system, the characteristic dynamical time therefore is $t_{\rm dyn}\sim A/\Gamma_{\rm tot}$, whereas the collisional relaxation time generated by discreteness effects scales as $ \tau_{\rm coll}\sim N_v\,t_{\rm dyn}$. Hence,
\begin{equation} 
\frac{\tau_{\rm coll}}{t_{\rm dyn}} \sim N_v , \end{equation} 
which diverges in the limit $N_v\to\infty$. The build-up of correlations is therefore suppressed, the collision operator $\mathcal C[f]$ vanishes, and the dynamics becomes exactly collisionless. For times intermediate between the fast microscopic fluctuations and the extremely slow secular relaxation, $t_{\rm dyn}\lesssim t\ll \tau_{\rm coll}$, the evolution of the vortex gas is therefore by the self-consistent mean field alone, which isolates the collisionless Vlasov regime in direct analogy with the framework used in collisionless plasmas and stellar dynamics \cite{nicholson,binney} and provides the foundation necessary to explore wave-vortex interactions and collisionless transport without the blurring effects
of discrete collisional diffusion. 

Neglecting the correlation term therefore leads to a kinetic equation of the Vlasov type,
\begin{equation}
\frac{\partial f}{\partial t}+\{f,\Psi\}=0,
\label{eq:vlasov2}
\end{equation}
self-consistently coupled to
\begin{equation}
\nabla^2\Psi=-\Gamma f.
\label{eq:poisson2}
\end{equation}
These equations constitute the fundamental kinetic description of a
collisionless point-vortex gas.

\section{Collective excitations in a vortex gas}
\label{sec:rossby}

We consider an axisymmetric equilibrium vortex distribution
\begin{equation}
f({\bf r},t)=f_0(r), \qquad \Psi({\bf r},t)=\Psi_0(r),
\label{4.1}
\end{equation}
which is an exact stationary solution of the Vlasov equation, with corresponding angular velocity
\begin{equation}
\Omega(r) = -\frac{1}{ r} \frac{d\Psi_0}{dr}.
\label{4.2}
\end{equation}
Introducing small perturbations about this state, $f=f_0+\delta f$ and
$\Psi=\Psi_0+\delta\Psi$, and retaining only first-order terms, the
linearized Vlasov equation reads
\begin{equation}
\frac{\partial \delta f}{\partial t} +
\{\delta f,\Psi_0\} + \{f_0,\delta\Psi\} = 0 .
\label{4.4}
\end{equation}
In polar coordinates, and for normal modes of the form
\begin{equation}
\delta f = f_m(r)\, e^{i(m\theta-\omega t)}, \qquad \delta\Psi = \psi_m(r)\, e^{i(m\theta-\omega t)},
\label{4.5}
\end{equation}
one obtains
\begin{equation}
(\omega-m\Omega) f_m = \frac{m}{r}\frac{df_0}{dr} \psi_m ,
\label{4.6}
\end{equation}
which, combined with the linearized Poisson equation
$\nabla^2\delta\Psi=-\Gamma\delta f$ following from
Eq.~\eqref{eq:poisson2},
\begin{equation}
\frac{1}{r} \frac{d}{dr} \left(r\frac{d\psi_m}{dr}\right) - \frac{m^2}{r^2}\psi_m =-\Gamma f_m ,
\label{4.8}
\end{equation}
allows $f_m$ to be eliminated and leads to the eigenvalue equation
\begin{equation}
\frac{1}{r} \frac{d}{dr}\left(r\frac{d\psi_m}{dr}\right) - \frac{m^2}{r^2}\psi_m + \frac{m\Gamma }{r} \frac{f_0'(r)} {\omega -m\Omega(r)}\psi_m = 0 . 
\label{4.9}
\end{equation}
Equation \eqref{4.9} determines the spectrum of collective excitations supported by an arbitrary vortex equilibrium, playing a role analogous to that of the Rayleigh equation of hydrodynamic stability and of the eigenvalue equations governing vortex Rossby waves
in geophysical fluids \cite{mcwilliams2003,montgomery1997}. The azimuthal number $m=1$ describes a rigid translation of the system (zero mode) and carries no restoring force, and thus $\omega_1=0$. Unlike ordinary gases, vortex gases support volume modes only when the equilibrium profile is either non-uniform or compact, satisfying $f_0'\neq 0$, since wave propagation here relies entirely on the self-consistent interaction between vortex-density gradients and the induced stream function. 

\subsection{Vortex Rossby-like waves}

A particularly interesting limit is obtained when the equilibrium varies on a scale $R$ much larger than the wavelength of the perturbation, $qR \gg 1$. In the spirit of the the WKB approximation, we consider such short-wavelength perturbations to be of the form 
\begin{equation}
\psi_m(r,\theta) \sim e^{i{\bf q}\cdot {\bf r}-i\omega t} = e^{i(q_\perp r +m\theta- \omega t)},
\label{4.10}
\end{equation}
so that, with a total wavenumber $q^2=q_\perp^2+m^2/r^2$,
Eq.~\eqref{4.9} yields the local dispersion relation
\begin{equation}
\omega = m\Omega(r) + \frac{m\Gamma }{r}\frac{f_0'(r)}{q^2}.
\label{4.13}
\end{equation}
Interestingly, the restoring force responsible for propagation is provided neither by compressibility, as
in sound waves, nor by an external Coriolis force, as in planetary Rossby waves \cite{holton_2004}, but originates directly in the gradients of the vortex density, a dynamical consequence of the phase-space restriction imposed by the commutation relations of Eq.~\eqref{2.5}. The quantity $f_0'(r)$ therefore plays the role of the planetary vorticity gradient of geophysical fluids. To render this explicit, we linearize the profile about a radius $R$, $f_0(r)\simeq f_0(R)+f_0'(R)(r-R)$, which is consistent with the short-wavelength ordering $qR\gg 1$ assumed above, and split the wavenumber in the $\beta$-plane fashion, $q^2=q_\parallel^2+q_\perp^2$
with $q_\parallel=m/R$. Equation~\eqref{4.13} then takes the more appealing form
\begin{equation}
\omega = U_0\, q_\parallel
- \beta\,\frac{q_\parallel}{q_\perp^{2}+q_\parallel^{2}} ,
\label{eq:Rossby}
\end{equation}
with $U_0=R\,\Omega(R)$ being the zonal flow speed and
$\beta=-\Gamma f_0'(R)$ the Rossby parameter, which is positive for the monotonically decreasing profiles of interest here. It is important to distinguish these excitations from the Tkachenko oscillations of an ordered vortex lattice \cite{tkachenko1966,tkachenko1969}. The present kinetic theory describes a disordered vortex gas interacting through its
self-consistent mean field,
whereas Tkachenko modes arise from the elastic degrees of freedom of a crystalline vortex array and correspond to transverse shear oscillations of the lattice \cite{baym2003,sonin2014}. The origin of
the restoring force is thus fundamentally different in the two cases, being the vortex-density gradient and the mean-field interaction for the kinetic modes considered here, and lattice elasticity for Tkachenko waves. In rapidly rotating superfluids exhibiting partial vortex ordering both mechanisms may nevertheless coexist, which suggests the possibility of a unified Rossby--Tkachenko description of collective vortex dynamics \cite{tercas_NJP,du2024}. In what follows, we examine two particularly relevant configurations, the first being a uniform vortex gas, whose spectrum can be determined exactly, while the second is a smooth vortex distribution, where critical layers emerge naturally and prepare the ground for the theory of vortical Landau damping developed in the next section.

\subsection{Global modes}

\begin{figure}[t!]
\includegraphics[width=\columnwidth]{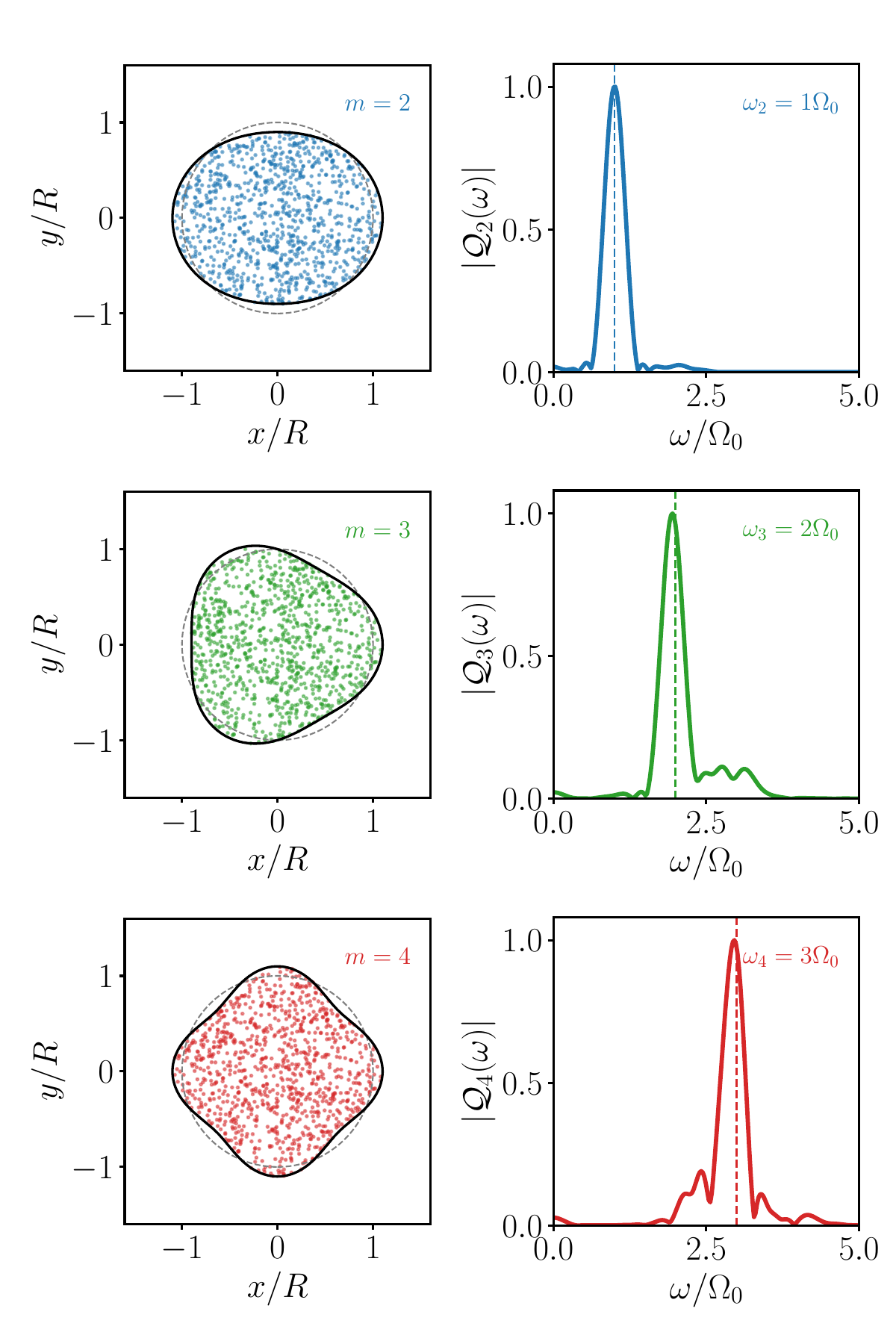}
\caption{(color online) Normal modes of a cold vortex gas composed of $N_v=1000$
vortices in a circular domain of radius $R$. Left column: snapshots of the spatial configurations for the modes $m=2$, $3$ and $4$. Right column: normalized Fourier transforms of the moments, $|\mathcal{Q}_m(\omega)|$, showing the resonance frequencies $\omega_m$ predicted by Eq.~\eqref{5.11}.}
\label{fig:patch}
\end{figure}

Consider a uniform vortex gas (Onsager patch) configuration,
\begin{equation}
f_0(r)=n_0\,\Theta(R-r),
\label{5.1}
\end{equation}
where $\Theta$ denotes the Heaviside function, $n_0$ the equilibrium vortex density and $R$ the patch radius. Uniform vortex patches occupy a distinguished place in the theory of two-dimensional turbulence and in point-vortex statistical mechanics, going back to Kirchhoff and Onsager \cite{onsager,kirchhoff}. Inside the patch the equilibrium stream function satisfies
\begin{equation}
\nabla^2\Psi_0=-\Gamma n_0,
\label{5.2}
\end{equation}
and regularity at the origin immediately gives
$\Psi_0'(r)=-\Gamma n_0 r/2$, so that
\begin{equation}
\Omega_0 = -\frac{1}{r} \frac{d\Psi_0}{dr} = \frac{\Gamma n_0}{2},
\label{5.4}
\end{equation}
the patch rotating as a rigid body with angular velocity $\Omega_0$, a manifestation of the {\it solid-body rotation limit} commonly invoked in the literature of rotating Bose-Einstein condensates
\cite{engels2003,simula2004,fetter2009,correggi2013}. For the equilibrium profile in Eq. \eqref{5.1}, the density gradient appearing in Eq.~\eqref{4.9} becomes $f_0'(r)=-n_0\delta(r-R)$, so that the
eigenvalue problem reduces to Laplace's equation everywhere except at
the boundary,
\begin{equation}
\frac{1}{r}\frac{d}{dr}\left(r\frac{d\psi_m}{dr}\right) -\frac{m^2}{r^2}\psi_m=0.
\label{5.6}
\end{equation}
Regularity at the origin and decay at infinity yield
$\psi_{\rm in}=Ar^m$ and $\psi_{\rm out}=Br^{-m}$ inside and outside the patch respectively, while continuity of the stream function at $r=R$ implies $B=AR^{2m}$. Integrating Eq.~\eqref{4.9} across the singular layer at $r=R$, we obtain the jump condition
\begin{equation}
\left[r\frac{d\psi_m}{dr}\right]_{R^-}^{R^+}=\frac{\Gamma m n_0}{\omega-m\Omega_0}\psi_m(R), 
\label{5.9}
\end{equation}
which, together with Eq.~\eqref{5.4}, yields
\begin{equation}
\omega_m = (m-1)\Omega_0,
\label{5.11}
\end{equation}
with $\Omega_0=\Gamma n_0/2$ being the characteristic {\it vortical frequency}, which plays a similar role as the plasma frequency of a system with long-range interactions \cite{nicholson}. This simple spectrum contains a zero mode,
$\omega_1=0$, associated with rigid translations of the entire patch. Higher modes correspond to collective shape oscillations: $m=2$ describes an elliptical deformation, $m=3$ a triangular distortion, and successive values of $m$ generate increasingly complex multipolar oscillations, in close relation to the classical Kelvin--Kirchhoff oscillations \cite{lamb,saffman}.
To corroborate these findings numerically, we distribute a collection of vortices in a circular patch of radius $R$ and evolve them in time according to Eq.~\eqref{2.2}, extracting the normal-mode dynamics from the multipolar moments
\begin{equation}
\mathcal{Q}_m(t)=\frac{1}{N_v}\sum_{\alpha=1}^{N_v}\ z_\alpha(t) ^m, \qquad z_\alpha =x_\alpha+i y_\alpha,
\label{5.10}
\end{equation}
which capture the oscillations of the $m$th mode. The results are shown
in Fig.~\ref{fig:patch}.

A second and physically richer situation arises for smooth profiles. As
a representative example we take the Gaussian vortex profile
\begin{equation}
f_0(r)=n_0\exp\!\left(-\frac{r^2}{a^2}\right),
\label{5.12}
\end{equation}
where $a$ defines a characteristic radius, and the resulting angular velocity profile follows is
\begin{equation}
\Omega(r)=\Omega_0 \frac{a^2}{r^2}\left(1-e^{-r^2/a^2}\right).
\label{5.15}
\end{equation}
Near the origin, $r\ll a$, one has $\Omega(r)\simeq\Omega_0$, so that
the gas behaves locally as a rigidly rotating core, whereas at large
radii $\Omega(r)\simeq\Omega_0a^2/r^2$ and differential rotation
becomes increasingly important. In contrast with the Onsager patch, the
angular velocity now varies continuously across the distribution and the denominator in Eq.~\eqref{4.9} may consequently vanish at some radius $r=r_c$ such that
\begin{equation}
\omega=m\Omega(r_c),
\label{5.19}
\end{equation}
the point at which the collective mode becomes resonant with the local
rotation. The radius $r_c$ thus defines a critical layer, analogous to the resonant surface encountered in hydrodynamic stability theory, in the theory of vortex Rossby waves and in collisionless plasma physics \cite{montgomery1997,mcwilliams2003}. Smooth vortex distributions carry a continuous family of local frequencies $m\Omega(r)$, so that discrete global modes coexist with a continuum of
resonant vortex motions. This emergent stratification constitutes the most important qualitative difference between sharp and smooth vortex equilibria, and the layers it produces provide a mechanism through which global oscillations can exchange energy and angular momentum with localized subsets of vortices, giving rise, as we show in the next section, to a vortex analogue of Landau damping.

\section{Vortical Landau damping}
\label{sec:landau}

One of the most remarkable consequences of the eigenvalue equation \eqref{4.9} is the appearance of critical layers satisfying the resonance condition \eqref{5.19}, a feature that suggests a damping mechanism closely analogous to Landau damping in collisionless plasmas \cite{landau1946}. In the plasma problem, particles of speed $v$ that are resonant with a mode of wavevector $q$ satisfy $\omega=qv$ and energy is transferred from the collective wave to those particles moving at the phase velocity of the perturbation. To
make the analogy explicit, it is useful to introduce the action-like variable,
\begin{equation}
J=\frac{r^2}{2},
\label{6.2}
\end{equation}
in terms of which the equilibrium quantities may be regarded as
functions of a single variable,
\begin{equation}
f_0=f_0(J),
\qquad
\Omega=\Omega(J).
\label{6.3}
\end{equation}
In these terms, Eq.~\eqref{4.6} gives the perturbed vortex density as
\begin{equation}
f_m= m\,\frac{f_0'(J)}{\omega-m\Omega(J)}\,\psi_m(J),
\label{6.4}
\end{equation}
which is formally identical to the response function of collisionless plasma theory, with the resonant denominator $\omega-qv$ replaced by $\omega-m\Omega(J)$, so that the resonant radius becomes the vortex counterpart of the resonant velocity of the classical Landau problem. As we are about to see, however, the non-canonical geometry carries substantive modifications.

\subsection{The dielectric operator and the damping rate}
\label{sec:dielectric}

Closing Eq.~\eqref{6.4} with the Poisson relation \eqref{eq:poisson2}, written in action variables with the kernel $\Lambda_m(J,J')$ derived in Appendix~\ref{app:lambda_m}, gives a homogeneous Fredholm equation of the second kind,
\begin{equation}
\psi_m(J)=m\int_0^\infty\!\! dJ'\,\Lambda_m(J,J')\,
\frac{f_0'(J')}{\omega-m\Omega(J')}\,\psi_m(J').
\label{eq:Fredholm}
\end{equation}
What Eq.~\eqref{eq:Fredholm} defines is a dielectric {\it operator},
\begin{equation}
\hat\varepsilon_m(\omega)=\hat{\mathbb I}-\hat K_m(\omega),
\label{eq:diel_operator}
\end{equation}
with $\hat K_m(\omega;J,J')=m\,\Lambda_m(J,J')
f_0'(J')/[\omega-m\Omega(J')]$, and the collective modes are those frequencies at which $\hat\varepsilon_m(\omega)$ becomes singular,
\begin{equation}
\lambda_m(\omega)\equiv \frac{\langle\chi_m|\hat K_m(\omega)|\psi_m\rangle}
{\langle\chi_m|\psi_m\rangle} =1,
\label{eq:solvability}
\end{equation}
where $\psi_m$ and $\chi_m$ are the right and left eigenvectors of $\hat K_m$ belonging to the eigenvalue branch $\lambda_m$. Since $\Lambda_m$ is symmetric, the left eigenvector is determined by the right one,
\begin{equation}
\chi_m(J)=\frac{f_0'(J)}{\omega-m\Omega(J)}\,\psi_m(J)=\frac{1}{m}f_m(J),
\label{eq:left_eigen}
\end{equation}
so that the adjoint problem is solved by the perturbed density itself. Notice that a local dielectric function may be recovered in the WKB limit, in which the radial wavenumber $q$ is large compared with the gradient scale and Eq.~\eqref{4.9} reduces to the local dispersion relation discussed in Sec. ~\ref{sec:rossby}
\begin{equation}
\varepsilon_m(q,\omega)\simeq 1-\frac{m\Gamma}{q^2}\,
\frac{f_0'(r)}{\omega-m\Omega(r)}=0.
\label{6.6}
\end{equation}
Resonant wave-particle processes enter through the analytic continuation of Eq.~\eqref{eq:Fredholm} from the upper half plane, $\gamma \equiv {\rm Im}\,\omega>0$, via the Landau prescription, which for an equilibrium with $\Omega'<0$ reads
\begin{equation}
\frac{1}{\omega-m\Omega}\;\longrightarrow\;
{\cal P}\frac{1}{\omega-m\Omega}-i\pi\delta(\omega-m\Omega).
\label{6.10}
\end{equation}
The dielectric operator splits as
$\hat K_m=\hat K^{\cal P}_m+i\hat K^{\,\rm res}_m$, with
\begin{align}
K^{\cal P}_m(J,J')&=m\,\Lambda_m(J,J')\,{\cal P}
\frac{f_0'(J')}{\omega-m\Omega(J')}, \nonumber \\[5 pt]
K^{\,\rm res}_m(J,J')&=-\pi\,\Lambda_m(J,J')\,
\frac{f_0'(J')\,\delta(J'-J_c)}{|\Omega'(J_c)|},
\label{eq:split}
\end{align}
where $J_c$ denotes the action of the critical layer,
$\omega_r=m\Omega(J_c)$. Writing $\omega=\omega_r+i\gamma$ with
$|\gamma|\ll\omega_r$, expanding Eq.~\eqref{eq:solvability} to first order and invoking the Hellmann--Feynman theorem \cite{Epstein_1966}, $\partial_\omega\lambda_m=\langle\chi_m|\partial_\omega\hat K_m|
\psi_m\rangle/\langle\chi_m|\psi_m\rangle$, the normalization
$\langle\chi_m|\psi_m\rangle$ cancels and one obtains the {\it vortical Landau damping} rate
\begin{align}
\gamma &=-\frac{\langle\chi_m|\hat K^{\,\rm res}_m|\psi_m\rangle}
{\langle\chi_m|\partial_\omega \hat K_m|\psi_m\rangle}
\nonumber \\[5 pt]
&=-\,\frac{\pi\,f_0'(J_c)\,\psi_m^2(J_c)}
{m\,|\Omega'(J_c)|\;\;{\displaystyle {\cal P}\!\!\int_0^\infty}\! dJ\,
{\displaystyle \frac{f_0'(J)\,\psi_m^2(J)}{[\omega_r-m\Omega(J)]^2}}} .
\label{6.17}
\end{align}
Equation \eqref{6.17} is the central result of this section and some comments are due. Being a ratio of two resonance-weighted moments of the same equilibrium profile, it is
invariant under rescaling of $\psi_m$. Also, the numerator contains the squared mode amplitude evaluated at the critical layer, $\psi_m^2(J_c)$, which plays here the role that the electrostatic energy density $|E_k|^2$ plays at the resonant velocity in the plasma problem \cite{nicholson}. Finally, since $\gamma<0$ whenever $f_0'<0$, monotonically decreasing equilibria are damped, in agreement with
Rayleigh's criterion \cite{rayleigh_1880, briggs1970, schecter2000}.

\subsection{Damped quasi-modes of a finite-edge patch}
\label{sec:finite_edge}

Equation \eqref{6.17} assumes the existence of a weakly damped discrete mode, $|\gamma|\ll\omega_r$, whose natural setting is a vortex patch with an edge smoothed over a finite width, since such an equilibrium supports a quasi-mode at $\omega_r\simeq(m-1)\Omega_0$ while possessing a genuine critical layer. Here, the edge pins the real frequency such that the jump condition \eqref{5.9} survives smoothing and fixing $\omega_r$ up to corrections of the order of the edge width. In the thin-edge limit, the
eigenfunctions are given as $\psi_m\sim (r/R)^m$ for $r<R$ and $\psi_m\sim (R/r)^m$ for $r>R$, and the principal-value integral in Eq.~\eqref{6.17} is dominated by the edge, where it reduces to $-n_0/\Omega_0^2$, so that Eq.~\eqref{6.17} takes the closed form
\begin{equation}
\gamma \simeq\frac{\pi\Omega_0^2}{m\,n_0}\,
\left(\frac{m-1}{m}\right)^{m}\,
\frac{f_0'(J_c)}{|\Omega'(J_c)|},
\label{6.21}
\end{equation}
with $J_c = J_R\,m/(m-1)$, in which the geometric factor $((m-1)/m)^m=|\psi_m(J_c)/\psi_m(J_R)|^2$ is precisely the suppression of the amplitude of the mode discussed above. Since the critical layer
sits {\it outside} the patch, $r_c>R$, the gradient $f_0'(J_c)$ samples the tail of the distribution, so that the damping is exponentially small in the edge width and is completely controlled by how much vorticity gradient survives at $r_c$. This is the vortex counterpart of the algebraically small damping of the diocotron electron columns 
\cite{briggs1970,schecter2000,bassom1998,balmforth2001}.

\subsection{Smooth profiles and the Landau pole}
\label{sec:smooth}

For a smooth monotonic equilibrium such as the Gaussian profile of Eq.~\eqref{5.12}, the continuum $m\Omega(J)$ covers the whole interval
$[0,m\Omega_0]$ and the condition \eqref{eq:solvability} admits no root on the real axis. What replaces the discrete eigenvalue
is a pole of the continued resolvent. Defining $x=2J/a^2=r^2/a^2$, the linearized problem \eqref{eq:Fredholm} reads
\begin{equation}
\omega f_m(x)=m\,\Omega(x)\,f_m(x)
\;\underbrace{-\,e^{-x}\!\int_0^\infty\!\!
\left(\frac{x_<}{x_>}\right)^{m/2}\!\!f_m(x')\,dx'}_{\textstyle \mathbb{K}_m},
\label{eq:eigen_x}
\end{equation}
with $\Omega(x)=(1-e^{-x})/x$, and separates the two competing agents:
the shear $m\Omega(x) f_m(x)$, which spreads the perturbation in frequency,
and the induced stream function $\mathbb{K}_m$, which holds it together. To get some numerical insight, we discretize Eq.~\eqref{eq:eigen_x} on a Gauss--Legendre grid of $M$
nodes over $x\in[0,x_{\max}]$ and rotate the contour as
$x=s\,e^{i\theta}$. Since $\Omega'<0$, the rotation swings the continuum down to $\arg\omega=-\theta$, uncovering the second sheet, while a pole of the resolvent stays put; poles are therefore identified as eigenvalues common to several $\theta$. Cross-checking $\theta=0.70$, $0.95$ and $1.20$ at $M=1200$--$2200$ and $x_{\max}=100$--$180$ leaves exactly one such eigenvalue,
\begin{equation}
\frac{\omega_2}{\Omega_0}=0.44497-0.15878\,i ,
\qquad
\frac{r_c}{a}=2.108 ,
\label{eq:pole_gaussian}
\end{equation}
stable to five figures. For $m\ge3$ there is none: every candidate
belongs to the rotated continuum and moves with $\theta$. This indicates that Landau
damping of a smooth vortex gas is limited to the quadrupole mode $m=2$, and such a selection rule follows from Eq.~\eqref{eq:eigen_x} without
reference to the pole. To render this clear, let $\lambda_m$ be the largest eigenvalue of $\mathbb{K}_m$ and $v_m(x)$ its eigenvector, which is the radial shape the induced field would impose on its own, and let $\sigma_m$ be the rms spread of $m\Omega(x)$ weighted by $|v_m|^2$, which is the
frequency detuning the shear imposes across that shape. A collective
mode exists only if the first can bridge the second,
\begin{equation}
\mathcal{R}_m\equiv\frac{|\lambda_m|}{\sigma_m}>1 .
\label{eq:criterion}
\end{equation}
For the Gaussian profile, $\mathcal{R}_m=1.78$, $0.98$, $0.63$, $0.45$ and $0.34$ at $m=2,\dots,6$, and
\begin{equation}
\mathcal{R}_m\simeq\mathcal{R}_2\left(\frac{2}{m}\right)^{3/2}
\label{eq:criterion_law}
\end{equation}
to better than one percent up to $m=8$, so that Eq.~\eqref{eq:criterion} fails beyond $m^{*}=2\mathcal{R}_2^{2/3}=2.94$. The exponent splits as $|\lambda_m|\propto m^{-3/4}$ against $\sigma_m\propto m^{+3/4}$: the resonance $\omega=m\Omega$ turns a given spread in $\Omega$ into an $m$ times larger spread in frequency, while the $m$th mode of the logarithmic interaction weakens through both its prefactor $\Gamma/2m$ and its range $(r_</r_>)^m$. The quadrupole clears the threshold by a factor $1.8$, the sextupole misses it by two percent, and everything above is torn apart by differential rotation
before the induced field can organize it. The marginality at $m=3$ is particularly interesting, meaning that  it behaves neither as a mode nor as free streaming.
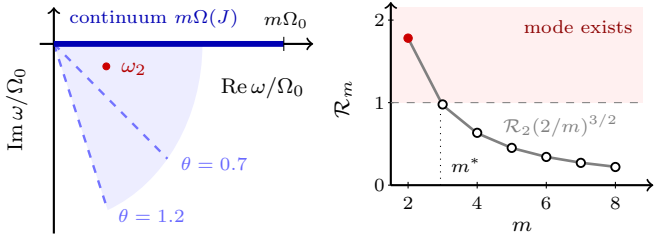
\begin{figure}[t!]
\centering
\begin{tikzpicture}[font=\large]
\pgfmathsetmacro{\ua}{1.56}
\pgfmathsetmacro{\ub}{0.7*\ua}
\pgfmathsetmacro{\ubx}{0.293*\ua}
\pgfmathsetmacro{\shiftb}{2.20*\ua+0.34}
\pgfmathsetmacro{\yshiftb}{(0.26*\ua-1.34*\ua)/2-(2.22*\ub-0.34*\ub)/2}
\begin{scope}[x=\ua cm,y=1.2*\ua cm]
\def\R{1.26}
\fill[blue!7] (0,0) -- (\R,0) arc (0:-68.75:\R) -- cycle;
\draw[line width=0.9pt,blue!55,dashed] (0,0) -- (-40.11:\R);
\draw[line width=0.9pt,blue!55,dashed] (0,0) -- (-68.75:\R);
\node[blue!60,font=\scriptsize,anchor=west] at (-40.11:1.31) {$\theta=0.7$};
\node[blue!60,font=\scriptsize,anchor=west] at (-68.75:1.30) {$\theta=1.2$};
\draw[->,thick] (-0.12,0) -- (2.20,0);
\draw[->,thick] (0,-1.34) -- (0,0.26);
\node[font=\footnotesize,anchor=north east] at (2.20,-0.17) {$\operatorname{Re}\omega/\Omega_0$};
\node[font=\footnotesize,rotate=90,anchor=south] at (-0.15,-0.46) {$\operatorname{Im}\omega/\Omega_0$};
\draw[line width=2.2pt,blue!70!black] (0,0) -- (1.95,0);
\draw (1.95,0.05) -- (1.95,-0.05);
\node[font=\scriptsize,anchor=south] at (1.97,0.05) {$m\Omega_0$};
\node[blue!70!black,font=\scriptsize,anchor=south west] at (0.06,0.05) {continuum $m\Omega(J)$};
\filldraw[red!80!black] (0.4450,-0.1588) circle (0.045cm);
\node[red!70!black,font=\footnotesize,anchor=west] at (0.50,-0.175) {$\omega_2$};
\end{scope}
\begin{scope}[xshift=\shiftb cm,yshift=\yshiftb cm,x=\ubx cm,y=\ub cm]
\fill[red!6] (1.5,1) rectangle (8.8,2.16);
\draw[->,thick] (1.5,0) -- (9.1,0);
\draw[->,thick] (1.5,0) -- (1.5,2.22);
\node[font=\footnotesize,anchor=north] at (5.3,-0.34) {$m$};
\node[font=\footnotesize,rotate=90,anchor=south] at (0.62,1.05) {$\mathcal{R}_m$};
\foreach \m in {2,4,6,8} \draw (\m,0.04) -- (\m,-0.04) node[below,font=\scriptsize] {\m};
\foreach \v in {0,1,2} \draw (1.42,\v) -- (1.58,\v) node[left,font=\scriptsize] {\v};
\draw[dashed,gray] (1.5,1) -- (8.8,1);
\node[font=\scriptsize,anchor=north east,text=red!55!black] at (8.75,2.10) {mode exists};
\draw[line width=1.0pt,gray] plot coordinates
  {(2,1.7809) (3,0.9694) (4,0.6296) (5,0.4505) (6,0.3427) (7,0.2720) (8,0.2226)};
\node[font=\scriptsize,anchor=west,text=gray] at (4.5,0.72) {$\mathcal{R}_2(2/m)^{3/2}$};
\draw[dotted] (2.938,0) -- (2.938,1);
\node[font=\scriptsize,anchor=south west] at (2.99,0.05) {$m^{*}$};
\filldraw[red!80!black] (2,1.7809) circle (0.055cm);
\foreach \m/\v in {3/0.9782, 4/0.6343, 5/0.4519, 6/0.3421, 7/0.2702, 8/0.2202}
  \filldraw[draw=black,fill=white,line width=0.7pt] (\m,\v) circle (0.055cm);
\end{scope}
\end{tikzpicture}
\caption{(color online) {\bf Contour rotation and the search for Landau poles}. Left panel: under $J\to Je^{i\theta}$, the continuum $m\Omega(J)$ aparts from the real axis $[0,m\Omega_0]$ and swings to $\arg\omega=-\theta$, uncovering the shaded sector of the second sheet. The single $\theta$-stable (i.e. fixed) eigenvalue is the $m=2$ pole of Eq.~\eqref{eq:pole_gaussian} (filled circle). For $m\ge3$, every candidate moves with $\theta$ and belongs to the rotated continuum. Right panel: The existence criterion in Eq. \eqref{eq:criterion}, collective frequency scale over shear detuning across the mode's own radial shape, against
Eq.~\eqref{eq:criterion_law} (line). It crosses unity at $m^{*}=2.94$, leaving $m=2$ above threshold and $m=3$ marginal by two percent.}
\label{fig:rotation}
\end{figure}
Three aspects deserve attention. First, the critical layer is much farther out than suggested by the uniform-vortex resonance. Equation~\eqref{eq:pole_gaussian} gives $\omega_r/m=0.22\,\Omega_0$, whereas $\omega_r=(m-1)\Omega_0$ would give $0.5\,\Omega_0$ and $r_c=1.26\,a$. Because $f_0'$ is $17$ times smaller at the actual resonance, Eq.~\eqref{6.21} substantially overestimates the damping rate when the uniform result is used. Second, the quadrupole is not weakly damped: $|\gamma|/\omega_r=0.357$. Equation~\eqref{6.21} identifies the damping mechanism, but not its magnitude. Third, the mode is collective. Suppressing the induced field reduces the amplitude at $\Omega_0t=15$ by a factor $9.1$, whereas the corresponding factors for $m=3$ and $m=4$ are only $0.80$ and $0.92$. In contrast to the finite-edge patch, where the edge fixes the mode frequency, a smooth vortex requires global self-consistency to sustain the quadrupole.

\begin{figure}[t!]
\centering
\includegraphics[width=\columnwidth]{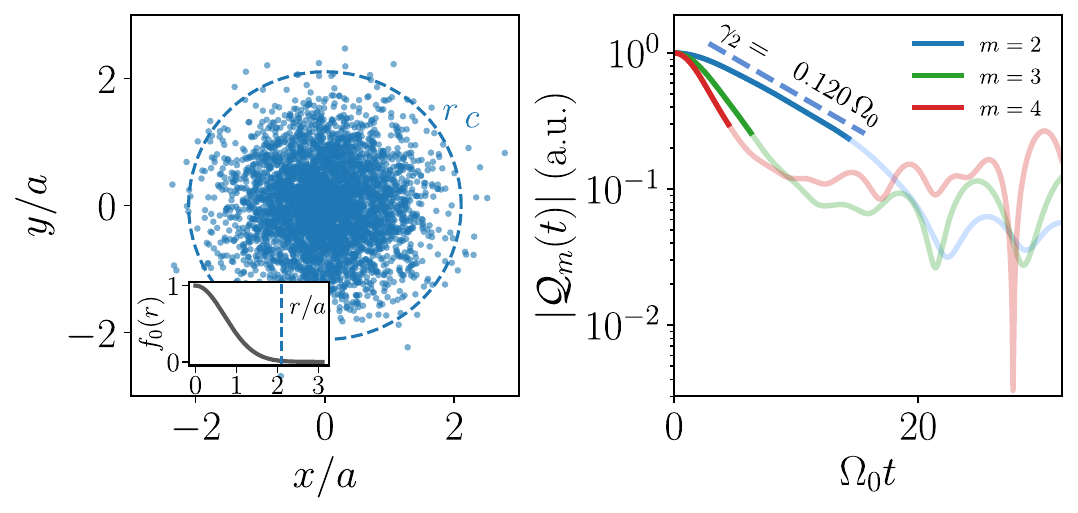}
\caption{(color online) {\bf Landau damping of a smooth vortex gas.}
Left: snapshot of the Gaussian equilibrium, with the critical layer of
the quadrupole, $r_c=2.11\,a$ from Eq.~\eqref{eq:pole_gaussian}, as a
dashed circle; the inset shows $f_0$ against $r/a$ with the same radius
marked, far out on the skirt where the vorticity gradient has fallen by
more than an order of magnitude. Right: multipole amplitude
$|\mathcal{Q}_m(t)|$ for $m=2,3,4$, each curve drawn solid while it
stands above its own discreteness floor and faded once it does not.
The perturbation is seeded only azimuthally, as
$f_m(r,0)\propto r\,e^{-r^2/a^2}$, the vortices being laid down on
equal-quantile rings with golden-angle offsets so that every multipole
starts at machine precision. The quadrupole decays as a single
exponential, dashed, at $\gamma_2=-0.120\,\Omega_0$ with
$\omega_2=0.432\,\Omega_0$, against $0.44497-0.15878\,i$ from
Eq.~\eqref{eq:pole_gaussian}, and its residual oscillation is at the
pole itself, neither at $(m-1)\Omega_0$ nor at $m\Omega_0$. The two
higher multipoles reach the floor within a few rotation times and admit
no exponential at all, which is Eq.~\eqref{eq:criterion} seen
directly.}
\label{fig:landau}
\end{figure}

These predictions are tested numerically in Fig.~\ref{fig:landau}. We integrate $N_v$ point vortices sampled from the Gaussian profile, perturbed so that $f_m(r,0)\propto r\,e^{-r^2/a^2}$, and measure $\mathcal{Q}_m(t)$. For $m=2$, the signal decays exponentially while remaining above the discreteness floor, with $\omega/\Omega_0\simeq0.43$ and $\gamma/\Omega_0\simeq-0.12$, in good agreement with Eq.~\eqref{eq:pole_gaussian}. The residual oscillation occurs at the quasi-mode frequency itself, not at $(m-1)\Omega_0$ or $m\Omega_0$. The remaining discrepancy is unsurprising: the quadrupole is strongly damped, with $|\gamma|/\omega_r\simeq0.36$, so the quasi-mode is not cleanly separated from the continuum.

\subsection{Phase mixing and the asymptotic tail} 
\label{sec:tail} 

The quasi-mode does not exhaust the response, and for $m\ge3$ no quasi-mode exists. At late times the dynamics is controlled by the branch cut $\omega\in[0,m\Omega_0]$. Near the upper endpoint, $\omega\to m\Omega_0$, we write $x_c\simeq2(1-\omega/m\Omega_0)$ and using $\Lambda_m(J,J_c)\propto J_c^{m/2}$, $\tilde\psi_m(J)\propto J^{m/2}$, and $|\Omega'(J_c)|\to\Omega_0/a^2$, the jump of $\tilde Q_m(\omega)$ $-$ the one-sided Fourier transform of $\mathcal{Q}_m$ defined as $ \mathcal{\tilde Q}_m(\omega)=\int_0^\infty e^{-i\omega t}\mathcal{Q}_m(t) dt$ $-$ across the cut, $\Delta \mathcal{\tilde Q}(\omega)=\mathcal{\tilde Q}(\omega+i0)-\mathcal{\tilde Q}(\omega-i0) $, scales as 
\begin{equation} \Delta\tilde{\mathcal{Q}}_m(\omega)\;\propto\; x_c^{\,m}\;+\;x_c^{\,(m+q)/2}, \qquad f_m(r,0)\sim r^{\,q}\;\;(r\to0). 
\label{eq:jump} 
\end{equation} 
For a regular perturbation, $q\le m$, so that the dominant exponent is $p=(m+q)/2$. Inverting the transform, yields 
\begin{equation} |\mathcal{Q}_m(t)|\;\longrightarrow\; \left(\frac{2}{m\Omega_0 t}\right)^{p+1}, \qquad p=\frac{m+q}{2}, \label{eq:algebraic} \end{equation} 
so that an analytic perturbation ($q=m$) decays as $t^{-(m+1)}$, whereas $q=1$ gives $t^{-(m+3)/2}$. Writing $\mathcal{Q}_m\propto\int_0^\infty dx\,x^p e^{-x} e^{-i\tau\phi(x)}$ with $\tau=m\Omega_0 t$ and $\phi(x)=(1-e^{-x})/x$, the phase has a saddle point at $x_s=\sqrt{\tau}\,e^{i\pi/4}$, corresponding to 
\begin{equation} 
\frac{r_*}{a}\simeq\left(m\Omega_0t\right)^{1/4}, \label{eq:saddle} 
\end{equation} 
and the steepest descent yields 
\begin{equation} \frac{|\mathcal{Q}_m(t)|}{|\mathcal{Q}_m(0)|}\;\simeq\; \frac{\sqrt{\pi}}{\Gamma_{\rm E}(p+1)}\, \left(m\Omega_0t\right)^{p/2+1/4} \exp\!\left[-\sqrt{2m\Omega_0t}\,\right], \label{eq:stretched} 
\end{equation} 
with $\Gamma_{\rm E}$ the Euler gamma function. Equation~\eqref{eq:stretched} reproduces direct numerical quadrature for $5\lesssim\Omega_0t\lesssim60$, whereas Eq.~\eqref{eq:algebraic} becomes accurate only after the crossover near $\Omega_0t\simeq10^2$. For $m=2$, the quasi-mode dominates throughout the accessible time window, as shown in Fig.~\ref{fig:rotation}. For $m\ge3$, Eq.~\eqref{eq:criterion} is not satisfied and the decay is entirely due to phase mixing, with Eq.~\eqref{eq:stretched} providing the relevant description. The algebraic regime of Eq.~\eqref{eq:algebraic} is, nevertheless, numerically inaccessible. By the time $\Omega_0t\sim10^2$, the signal has already fallen to $\sim10^{-5}$ of its initial value for $m=2$ and $\sim10^{-8}$ for $m=4$, while the estimator $\langle z^m\rangle$ carries a discreteness noise of order $N_v^{-1/2}$. Resolving the asymptotic tail would therefore require $N_v\sim10^{12}$ and $N_v\sim10^{18}$ vortices, respectively.

\section{Vortex--ring instability}
\label{sec:beam_vortex}

The dielectric formalism developed in the previous sections suggests a
vortex analogue of the celebrated beam--plasma instability problem. For that, we now examine a composite equilibrium consisting of a dense vortex patch surrounded by a thin annular
population, the patch providing the collective mode while the ring acts
as a resonant beam capable of exchanging energy with the global oscillation. We consider an axisymmetric equilibrium of the form
\begin{equation}
f_0(J) = n_0 \Theta(J_R-J) + n_b \delta(J-J_b),
\label{eq:ring_equil}
\end{equation}
where $J_R=R^2/2$ and $J_b=R_b^2/2$, with  $R_b>R$. The first term describes a uniform vortex patch of density $n_0$ and radius $R$,
whereas the second represents a narrow annular beam located at radius
$R_b$. Since $[\delta(J)]=L^{-2}$, the coefficient $n_b$ is
dimensionless, and the natural control parameters are the ratio of
vortex numbers and the ratio of radii,
\begin{equation}
\eta = \frac{n_b}{n_0 J_R}=\frac{N_{\rm ring}}{N_{\rm patch}},
\qquad \xi = \frac{R_b}{R},
\label{eq:ring_params}
\end{equation}
with $\eta$ playing the role of the beam-to-plasma density ratio. The
unperturbed angular frequency generated by the patch and ring together
follows from the enclosed circulation,
$\Omega(J)=(\Gamma/2J)\int_0^J f_0\,dJ'$, as
\begin{equation}
\Omega(J) =
\left\{
\begin{array}{ll}
\Omega_0, & J<J_R, \\[2mm]
\Omega_0\,J_R/J, & J_R<J<J_b, \\[2mm]
\Omega_0J_R(1+ \eta )/J, & J>J_b .
\end{array}
\right.
\label{eq:ring_profile}
\end{equation}
By the two-dimensional Gauss theorem, the ring produces no field in its own interior, so that the modes $(m-1)\Omega_0$ of the patch
remain unshifted to all orders in $\eta$; the patch, conversely, does shift the ring, and the ring shifts itself by one half of its own
circulation,
\begin{equation}
\Omega(J_b)=\frac{\Omega_0}{\xi^{2}}\left(1+\frac{\eta}{2}\right).
\label{eq:ring_mean}
\end{equation}
This self-field term is what renders
the thin-ring limit finite, and, as we shall see, it is also what quenches the instability. Note further that $f_0'$ changes
sign across the equilibrium, so that the necessary condition for instability is met, the sign change of $f_0'$ being precisely what
Rayleigh's inflection criterion requires \cite{drazinreid}.

The computation of the dispersion relation deserves some care. A ring of zero width cannot be inserted into Eq.~\eqref{eq:Fredholm}
directly, as it generates $\delta'(J-J_b)$ against a kernel
$\Lambda_m(J,J')$ that is not differentiable at $J=J'$. Regularizing
the ring to finite width and taking the thin limit at fixed circulation, the singular contributions cancel and the equilibrium is
left with three vorticity jumps, the patch edge and the two faces of
the sheet, so that Eq.~\eqref{eq:Fredholm} collapses onto a
$3\times3$ problem for the three surface amplitudes, as detailed in Appendix
\ref{app:thin_ring}. After regularization, the dispersion relation follows as
\begin{equation}
\left(1+\frac{\Omega_0}{D_0}\right)
\left(1-\frac{\omega_b^{2}}{D_-D_+}\right)
+\frac{\Omega_0\,g_m^{2}}{D_0D_-D_+}=0,
\label{eq:ring_cubic}
\end{equation}
where $D_0=\omega-m\Omega_0$ and $D_\pm=\omega-m\Omega(J_b^\pm)$ are
built on the patch frequency and on the two one-sided limits of
Eq.~\eqref{eq:ring_profile} at the ring, and
\begin{align}
\omega_b^{2}&=\frac{m\eta}{\xi^{4}}
\left[\,1-\xi^{-2(m-1)}-\frac{(m-1)\eta}{2}\,\right]\Omega_0^{2},
\label{eq:ring_wb}\\[2pt]
g_m^{2}&=\frac{m(m-1)\,\eta}{\xi^{2m+4}}
\left(\xi^{2}-1-\frac{\eta}{2}\right)\Omega_0^{2} .
\label{eq:ring_g}
\end{align}
Equation \eqref{eq:ring_cubic} is exact in $\eta$ within the thin-ring
model, is of degree three, one root for each vorticity jump, and has
been verified symbolically against the characteristic polynomial of the
three-interface problem for $m=2,3,4$. At $\eta=0$, it factorizes into
$\omega=(m-1)\Omega_0$, the modes of the patch, and a doubly degenerate passive ring at $\omega=m\Omega_0/\xi^{2}$.

Two features of Eq.~\eqref{eq:ring_cubic} distinguish the vortex
problem from its plasma counterpart, and both are visible in the ring
denominators. First, since the ring
carries a fraction $\eta$ of the patch circulation, the shear jumps
across it,
\begin{equation}
\Omega(J_b^-)=\frac{\Omega_0}{\xi^{2}},
\qquad
\Omega(J_b^+)=\frac{\Omega_0(1+\eta)}{\xi^{2}},
\label{eq:ring_faces}
\end{equation}
so that the two faces resonate at different frequencies and the ring
responds as a sheet rather than as a layer. Its mean orbital frequency
$\Omega(J_b)$ of Eq.~\eqref{eq:ring_mean} is the average of the two,
and it never enters Eq.~\eqref{eq:ring_cubic} on its own. The second aspect regards the physical content in $\omega_b$. By decoupling the ring from the patch, $g_m^{2}\to 0$, leaves the interior (patch) modes together with a ring doublet
\begin{equation}
\omega=m\Omega(J_b)\pm\sqrt{\omega_b^{2}+\frac{\eta^2}{4\xi^4}m^2\Omega_0^2},
\label{eq:doublet}
\end{equation}
the counterpart of the fast and slow beam modes of plasma physics
\cite{nicholson,briggs1970}. Since the slow branch carries negative energy \cite{cairns1979,sturrock1958}, what follows is the resonant coupling of a positive-energy mode of the bulk to a negative-energy mode of the beam. Note that $\omega_b$ is not an
orbital frequency, the capitals $\Omega_0$ and $\Omega(J)$ being
reserved for those, but the vortical counterpart of the beam plasma
frequency, with $\omega_b^{2}\propto\eta$ just as
$\omega_{pb}^{2}\propto N_b$. It is, however, a {\it dressed} one: as
shown in Appendix \ref{app:thin_ring}, the bare response of the ring to
its own induced field carries no $\xi^{-2(m-1)}$, and that factor
appears only once the patch is allowed to respond, the field the ring
raises on the patch edge acting back on the ring with the opposite
sign. The terms $-(m-1)\eta/2$ in
Eqs.~\eqref{eq:ring_wb}--\eqref{eq:ring_g} are the self-detuning
anticipated in Eq.~\eqref{eq:ring_mean}, and they enter both coefficients with the same weight.

\begin{figure}[t!]
\centering
\includegraphics[width=\columnwidth]{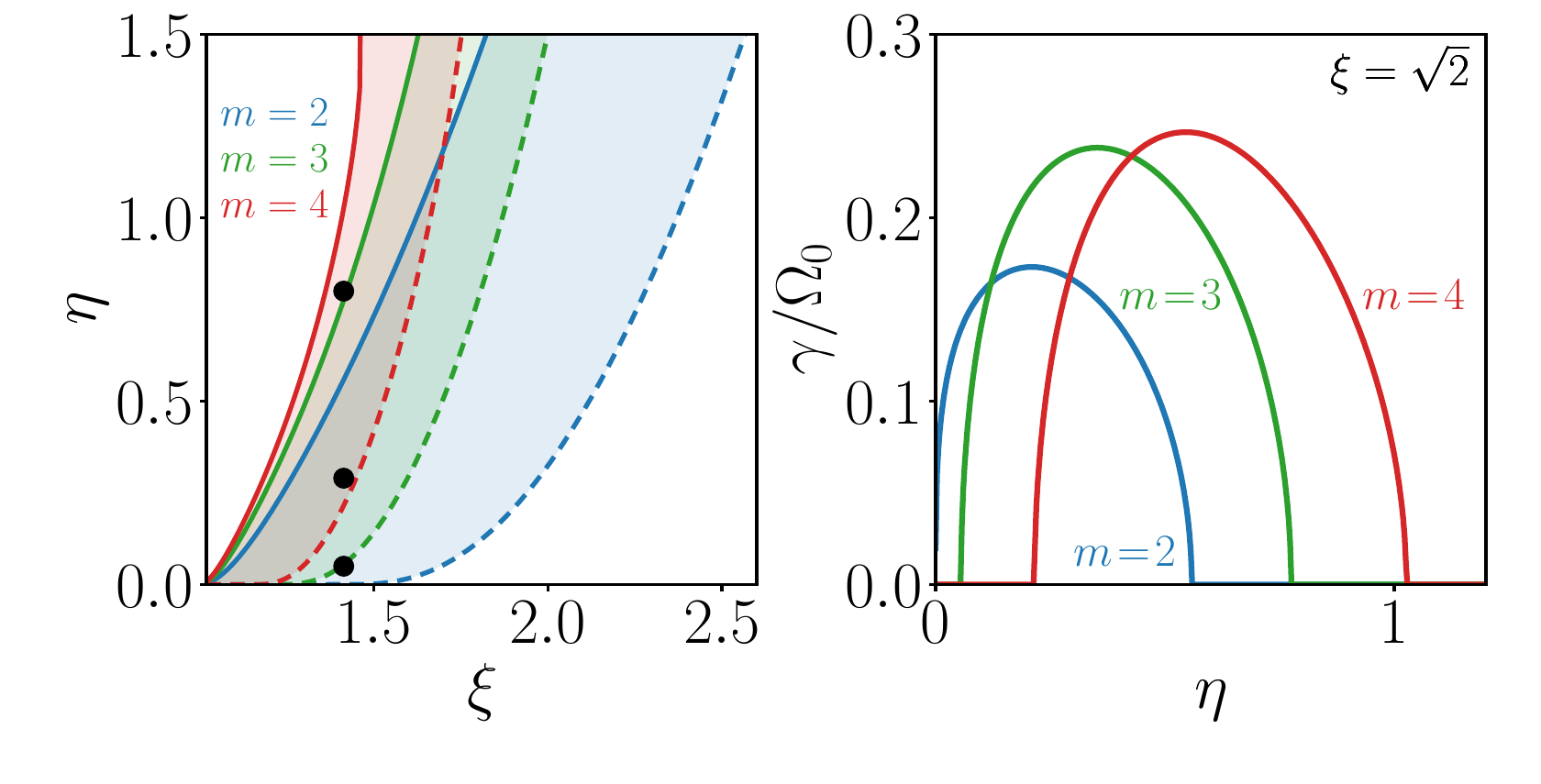}
\caption{(color online) {\bf Stability diagram of the patch--ring system}, from the dielectric determinant \eqref{eq:ring_cubic}. Left:
unstable region of each multipole in the $(\xi,\eta)$ plane, bounded by
its inner (solid) and outer (dashed) edge. As $\eta\to0$ every window
closes onto its own resonance $\xi_{\rm res}=\sqrt{m/(m-1)}$,
Eq.~\eqref{eq:xi_res}, the innermost belonging to the highest
multipole, whereas as $\eta$ grows the ring detunes itself through
Eqs.~\eqref{eq:ring_faces} and \eqref{eq:ring_wb} and the whole window
migrates outward. The windows therefore never close but move, so that a
mode held at fixed radius is quenched once the window has passed it,
which is the content of $\eta_c$, Table~\ref{tab:etac}. Black dots mark three
ring populations at $\xi=\sqrt2$, $\eta=0.05$, $0.30$ and $0.80$. Right: growth rate along that vertical line, each
multipole unstable over a finite interval of ring population and the
intervals ordered in $m$, so that raising $\eta$ at fixed radius
carries the resonance from one Kelvin level to the next.}
\label{fig:disp_numeric}
\end{figure}

\subsection{Weak ring limit}
\label{sec:weak_ring}

For $\eta\to0$ the ring modes degenerate onto $m\Omega(J_b)$ and the
instability reduces to a resonant collision with a single Kelvin level.
Setting $(m-1)\Omega_0=m\Omega(J_b)$ in Eq.~\eqref{eq:ring_mean} gives
the resonant radius
\begin{equation}
\xi_{\rm res}=\sqrt{\frac{m}{m-1}},
\label{eq:xi_res}
\end{equation}
always larger than unity and divergent for $m=1$, so that the
translational Goldstone mode, having no resonant partner, cannot be
destabilized. Here lies the sharpest difference with the plasma
problem. In a beam--plasma system the resonance $\omega_p=kv_b$ is met
by tuning the beam velocity, an independent parameter, whereas in a
vortex gas there is no independent momentum variable, the ``velocity''
of the ring being the frequency $\Omega(J_b)$ generated by the patch
itself. Resonance is therefore a purely geometric condition on $\xi$,
fixed by $m$ alone, and a direct consequence of the non-canonical
structure of Eq.~\eqref{2.7}, in which the plane is its own phase
space.

Write $\delta\equiv D_0+\Omega_0=\omega-(m-1)\Omega_0$ for the
departure from the Kelvin level. Because
$m\Omega(J_b^-)=m\Omega_0/\xi_{\rm res}^{2}=(m-1)\Omega_0$, the {\it
inner} face of the ring is exactly resonant there,
\begin{equation}
D_-=\delta
\qquad\text{at}\quad\xi=\xi_{\rm res},
\label{eq:inner_face}
\end{equation}
while the outer face is detuned from the Kelvin level by the full
circulation of the ring; equivalently, by Eq.~\eqref{eq:doublet}, the
mean displacement of the doublet equals the half-splitting $h$ of its
two faces. The resonance condition therefore does not place the ring
{\it as a whole} on the Kelvin level. It places one of its two faces
there, and the ring population controls how far the other has moved away, as we discuss below. Setting $\xi=\xi_{\rm res}$, Eq.~\eqref{eq:ring_cubic} reduces to
\begin{equation}
\delta^{3}-\omega_b^{2}\,\delta+\Omega_0\,g_m^{2}\simeq0 ,
\label{eq:cubic_weak}
\end{equation}
in which $\delta\sim\eta^{1/3}\Omega_0$, so that the linear term is
subdominant by $\eta^{1/3}$. The leading balance is therefore
$\delta^{3}\simeq-\Omega_0g_m^{2}$ and involves the coupling alone;
the screening of Eq.~\eqref{eq:ring_wb} is carried by the subdominant
term and does not enter. Using $(m-1)(\xi^{2}-1)=1$, which holds
identically at $\xi_{\rm res}$, the coupling itself collapses to
$g_m^{2}\to m\eta\,\Omega_0^{2}/\xi^{2m+4}$ as $\eta\to0$, and writing
$\gamma={\rm Im}(\delta)$,
\begin{equation}
\gamma\simeq\frac{\sqrt3}{2}\,(m\eta)^{1/3}
\left(\frac{m-1}{m}\right)^{(m+2)/3}\Omega_0 .
\label{eq:gamma_max}
\end{equation}
The most unstable mode therefore grows at $\gamma\propto\eta^{1/3}$,
exactly as in the reactive cold-beam instability of plasma physics \cite{nicholson}.

\subsection{Strong ring limit}
\label{sec:strong_ring}

At finite population, the isolated Krein collision of
Sec.~\ref{sec:weak_ring} is undone by three mechanisms: (i) the separation of the two faces, Eq.~\eqref{eq:ring_faces}, which lifts the degeneracy the collision requires; (ii) the shift of the ring
angular frequency, the ring feeling one half of its own circulation by Eq.~\eqref{eq:ring_mean}, so that the doublet center sits a distance
$\Delta\equiv m\,\delta\Omega(J_b)=(m-1)\eta\Omega_0/2$ above the Kelvin level at $\xi=\xi_{\rm res}$ and the resonance migrates outward; and (iii) the screening factor $1-\xi^{-2(m-1)}$ of Eq.~\eqref{eq:ring_wb}, which is geometric, survives at $\eta\to0$, and
multiplies the ring's response to its own field by $1/2$ at $m=2$ and by $1-e^{-1}$ at large $m$. The first two grow linearly in $\eta$, whereas the drive, entering as $(\Omega_0g_m^{2})^{1/3}$ through
Eq.~\eqref{eq:gamma_max}, rises only as $\eta^{1/3}$: the ratio $\Delta/\gamma$ runs from $0.05$--$0.09$ at $\eta=10^{-2}$ to $0.5$--$0.9$ at $\eta=0.3$ for $m=2$ to $4$. The instability is therefore bounded from above in $\eta$.

Keeping the ring at the unperturbed resonant radius,
Eq.~\eqref{eq:inner_face} yields $D_-D_+=\delta^{2}-2\Delta\delta$ identically, and Eq.~\eqref{eq:ring_cubic} becomes
\begin{equation}
\delta^{3}-2\Delta\,\delta^{2}-\omega_b^{2}\,\delta
+\Omega_0\,g_m^{2}=0,
\label{eq:cubic_res}
\end{equation}
with $\omega_b^{2}$ and $g_m^{2}$ of
Eqs.~\eqref{eq:ring_wb}--\eqref{eq:ring_g} evaluated at
$\xi_{\rm res}$. Setting $\Delta=0$, we recover
Eq.~\eqref{eq:cubic_weak}. The quadratic coefficient is exactly $-2\Delta$ and there is no term in $\Delta^{2}\delta$, which averaging the two faces would produce at order $\eta^{2}$. The instability terminates where two roots of Eq.~\eqref{eq:cubic_res} coalesce,
that is where its discriminant vanishes. Divided by its double root at $\eta=0$ that discriminant is a quartic in $\eta$ that does not admit useful reduction, and $\eta_c$ is collected in Table~\ref{tab:etac}. A closed
form exists for $m\gg1$: writing $\eta=y/m$, every coefficient of Eq.~\eqref{eq:cubic_res} converges, since $\xi_{\rm res}^{-2(m-1)}$ and
$\xi_{\rm res}^{-2m}$ both tend to $e^{-1}$, and the cubic becomes universal,
\begin{equation}
\delta^{3}-y\,\delta^{2}
-y\left(1-e^{-1}-\frac{y}{2}\right)\delta
+e^{-1}y\left(1-\frac{y}{2}\right)=0
\label{eq:cubic_univ}
\end{equation}
in units of $\Omega_0$. Its discriminant vanishes at $y_c=0.7631$, whence
\begin{equation}
\eta_c\simeq\frac{0.763}{m}
\qquad (m\gg1),
\label{eq:etac}
\end{equation}
the $1/m$ scaling of the geometric screening --- high multipoles couple
weakly to distant annuli and are quenched by correspondingly weaker
rings --- now carrying a definite coefficient in which $e$ is the only
constant to appear. Equation \eqref{eq:etac} is low by $6\%$ at $m=10$
and by $1\%$ at $m=60$, but by $32\%$ at $m=2$.

\begin{table}[t]
\label{tab:etac}
\begin{ruledtabular}
\begin{tabular}{ccccccc}
$m$ & $\xi_{\rm res}$ & $\eta_c$ & $m\,\eta_c$
    & $\gamma_{\rm peak}/\Omega_0$ & $\eta_\ast/\eta_c$
    & $\left(\tfrac{m}{m-1}\right)^{m}\tfrac{G_m}{\Omega_0}$ \\
\hline
2  & 1.4142 & 0.5584 & 1.117 & 0.1732 & 0.376 & 1.179 \\
3  & 1.2247 & 0.3223 & 0.967 & 0.2027 & 0.359 & 1.130 \\
4  & 1.1547 & 0.2267 & 0.907 & 0.2153 & 0.351 & 1.107 \\
6  & 1.0954 & 0.1422 & 0.853 & 0.2267 & 0.344 & 1.085 \\
10 & 1.0541 & 0.0815 & 0.815 & 0.2351 & 0.337 & 1.068 \\
\hline
$\infty$ & 1 & $0.763/m$ & 0.763 & 0.2468 & 0.330 & 1.044 \\
\end{tabular}
\end{ruledtabular}
\caption{Stability threshold of a ring held at $\xi_{\rm res}(m)$, from
the exact secular equation \eqref{eq:cubic_res}. The entries are roots
of its discriminant and are not approximated. The fourth column shows
the approach to the universal limit \eqref{eq:etac}; the last two give
the position of the maximum along the branch and the prefactor of the
square-root law \eqref{eq:sqrt_fold}, normalized to the geometric
coupling $[(m-1)/m]^{m}\Omega_0$.}
\end{table}

Since two roots merge at $\eta_c$, the complex pair detaches with the
square-root signature of a fold,
\begin{equation}
\gamma\simeq G_m\left(1-\frac{\eta}{\eta_c}\right)^{1/2},
\qquad
G_m\simeq\left(\frac{m-1}{m}\right)^{\!m}\Omega_0 ,
\label{eq:sqrt_fold}
\end{equation}
the prefactor being the geometric coupling itself, which the measured
$G_m$ exceeds by $18\%$ at $m=2$ and by $7\%$ at $m=10$. Together with
$\gamma\propto\eta^{1/3}$ of Eq.~\eqref{eq:gamma_max},
Eq.~\eqref{eq:sqrt_fold} brackets the branch from both ends: the growth
rate rises as the cube root of the ring population, turns over at
$\eta_\ast=0.376\,\eta_c$ for $m=2$, drifting to $0.330\,\eta_c$ as
$m\to\infty$, and vanishes as a square root at $\eta_c$. The peak rate
itself rises with $m$ and saturates at
$\gamma_{\rm peak}\simeq0.25\,\Omega_0$, the value set by
Eq.~\eqref{eq:cubic_univ}. A vortex ring is thus most destructive when
it carries about a third of the population that would stabilize it, and
no ring at any population drives a multipole faster than a quarter of
the rotation frequency of the patch.

Of the three mechanisms only the two detunings close the window, the
screening merely lowering the susceptibility at every $\eta$. Moreover
the unstable window in $\xi$ does not close but migrates outward,
following the drift of the resonance, so that a ring pinned at
$\xi_{\rm res}(m)$ is left behind by it while a ring free to sit
wherever the growth rate is largest would remain unstable throughout
the domain in which the thin-ring model holds. Table~\ref{tab:etac} is
therefore the threshold for a given multipole at a given radius, above
which the migrating window engages the neighboring Kelvin level and the
instability reappears at higher $m$.

\subsection{Simulation of the vortex-ring instability}
\label{sec:ring_sim}

\begin{figure*}[t!]
\includegraphics[width=\linewidth]{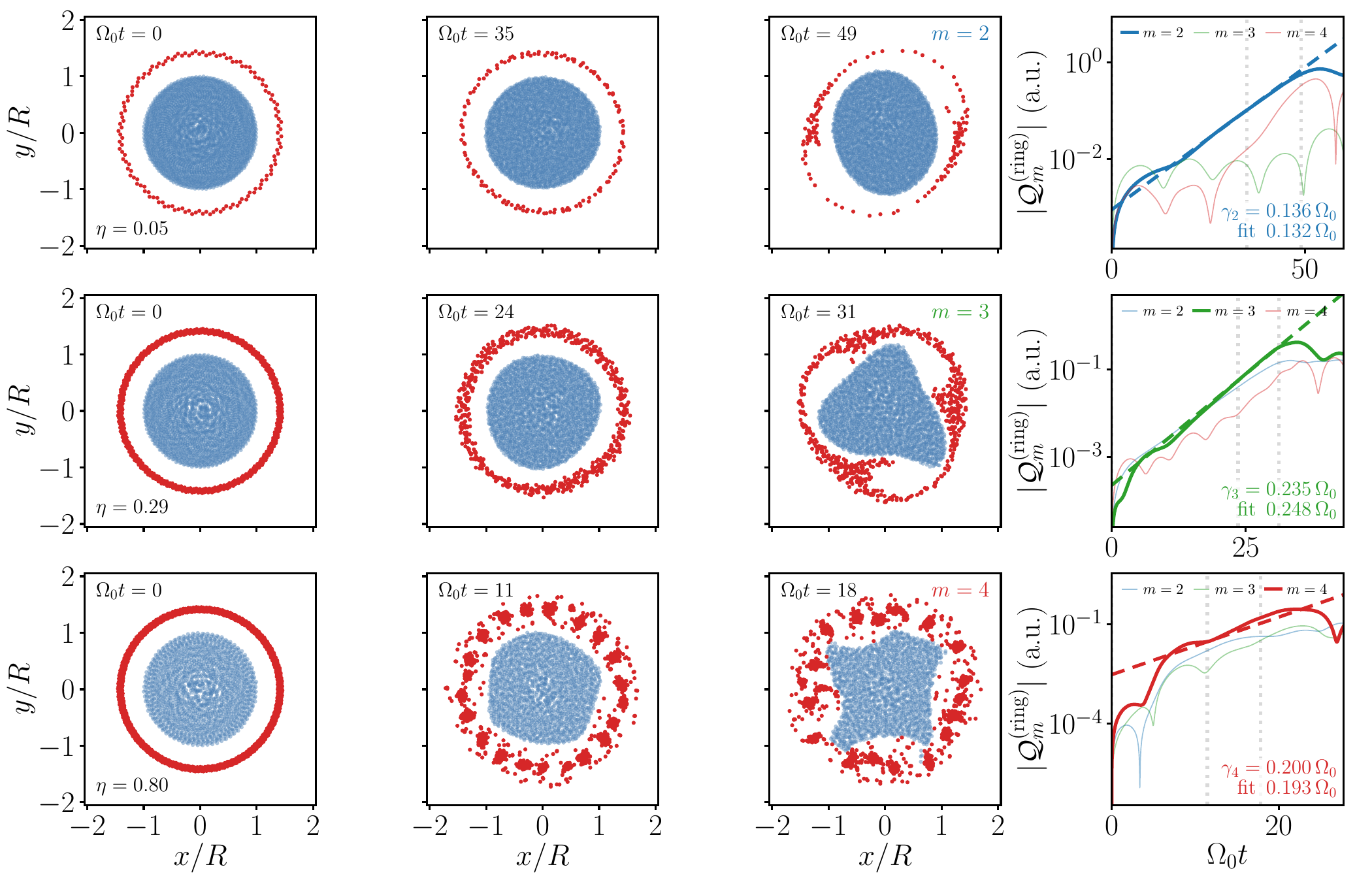}
\caption{(color online) {\bf Mode-selective vortex-ring instability.}
Uniform patch (blue) and thin ring (red) at fixed radius $\xi=\sqrt2$,
with the vortex population distributed between the two according to the
ring fraction $\eta$. From top to bottom, $\eta=0.05$, $0.29$ and
$0.80$. The first three columns show snapshots, at $t=0$ and at the two
times marked by the dotted vertical lines in the fourth column, which
displays the bunching parameter $|\mathcal Q_m^{\rm(ring)}|$ of
Eq.~\eqref{eq:bunching} for $m=2,3,4$, the dominant multipole being
highlighted by the heavier curve. Both populations are laid down so
that every azimuthal harmonic starts at machine precision, the patch on
equal-area shells and the ring equally spaced in angle, so that no mode
is seeded and the instability selects it spontaneously. Dashed lines
show exponential growth at the rates predicted by the dispersion
relation Eq.~\eqref{eq:ring_cubic}, $\gamma/\Omega_0=0.136$, $0.235$
and $0.200$ for the dominant $m=2$, $3$ and $4$ modes at the three
values of $\eta$ shown, against the measured $0.132$, $0.248$ and
$0.193$; only the intercept of each dashed line is taken from the data.
As the ring population increases, the resonance detunes from one Kelvin
level and engages the next, producing two, three and four azimuthal
clumps in succession. Each amplitude panel is cut at $1.25$ times its
own peak, the three rows saturating at different times.}
\label{fig:instability}
\end{figure*}

To corroborate the properties of both weak- and strong-ring limits, and
put in evidence the selective nature of the instability, we integrate
Eq.~\eqref{2.2} for a uniform patch and a ring held at $\xi=\sqrt2$,
with $N_v=3000$ vortices distributed between the two according to
$\eta$. Both are laid down so that every azimuthal harmonic starts at
machine precision, the patch on equal-area shells and the ring equally
spaced in angle: no mode is seeded, and what excites the spectrum is
the discreteness noise the patch regenerates as it mixes, which favours
no harmonic over another. The natural observable is the ring bunching
amplitude,
\begin{equation}
\mathcal{Q}^{\rm(ring)}_m(t)=\frac{1}{N_{\rm ring}}
\sum_{\alpha\in{\rm ring}} e^{\,i m\theta_\alpha(t)},
\label{eq:bunching}
\end{equation}
the direct counterpart of the beam bunching parameter of plasma
physics, and the quantity whose linear growth rate is $\gamma$. The
all-vortex moment of Eq.~\eqref{5.10} may be used instead, but it adds
the patch, whose Kelvin rotation beats against the ring and buries the
exponential under a large oscillating background.

The results are collected in Fig.~\ref{fig:instability}. At fixed ring
radius, increasing $\eta$ carries the resonance from one Kelvin level
to the next, so that the ring breaks into two, three and four clumps in
turn, which is the mode conversion of Sec.~\ref{sec:strong_ring} seen
directly in configuration space. The selection is unambiguous, and for
the lightest and heaviest rings it is not even a competition: at
$\xi=\sqrt2$ the windows of Eq.~\eqref{eq:ring_cubic} leave $m=2$ the
only unstable multipole for $\eta<0.056$ and $m=4$ the only one for
$\eta>0.775$, while at intermediate population all three are unstable
and the fastest is selected. The measured rates follow the prediction
with no adjustable parameter: $\gamma/\Omega_0=0.132$ against the
predicted $0.136$ for $m=2$ at $\eta=0.05$, $0.248$ against $0.235$ for
$m=3$ at $\eta=0.29$, and $0.193$ against $0.200$ for $m=4$ at
$\eta=0.80$, each fitted over the exponential stage alone.

\section{Conclusions} 

We have shown that a collisionless point-vortex gas behaves as a genuine Vlasov system. Starting from the non-canonical Hamiltonian structure of point-vortex dynamics, we derived a reduced Klimontovich equation and the corresponding Vlasov description directly on the physical plane, without introducing auxiliary momentum variables. The resulting theory supports the same hierarchy of collective phenomena familiar from other long-range interacting systems, including Rossby-like waves, global modes, critical layers, phase mixing, Landau damping, and beam-driven instability. For compact vortex patches, the spectrum reduces to the classical Kelvin--Kirchhoff modes. 

For smooth equilibria, differential rotation generates a continuous spectrum and critical layers satisfying $\omega=m\Omega(r_c)$. These resonances provide a collisionless channel for the transfer of energy and angular momentum from collective oscillations to localized vortex populations, producing a vortex analogue of Landau damping. More remarkably, differential rotation acts as a mode-selection mechanism. For the Gaussian equilibrium considered here, only the quadrupolar mode survives as a genuine quasi-mode, while all higher multipoles are destroyed by phase mixing before self-consistency can organize a global oscillation. The dielectric formalism developed here also reveals a close correspondence between vortex gases and collisionless plasmas. Critical layers play the role of resonant particle populations, the dielectric operator replaces the plasma dielectric function, and the quadrupolar quasi-mode appears as the analogue of a Landau pole. 

Likewise, the patch--ring configuration provides a direct counterpart of the beam--plasma instability, with the vortex ring acting as a resonant beam and the patch providing the collective mode. The resulting instability is mode-selective: as the ring population increases, the resonance migrates from one Kelvin level to the next, generating azimuthal clustering with a symmetry determined by the ring fraction. An important consequence of the kinetic picture is that neither mechanism acts uniformly across the gas. Landau resonance deposits energy and angular momentum into a narrow annulus around the critical layer, whereas the ring instability reorganizes an annular overdensity into a small number of discrete azimuthal clumps. In both cases, the rearrangement is localized in radius, structured in angle, collisionless, and operates on the dynamical timescale rather than on the secular timescale traditionally associated with vortex-gas kinetics. 

Whereas existing kinetic theories have largely focused on the slow collisional relaxation of the coarse-grained profile \cite{dubinoneil1988,chavanis2001,chavanis2008,chavanis2012}, what emerges here is fast, selective, and collective. These features suggest a possible connection with the dynamics of rotating neutron stars. The standard glitch picture involves the sudden transfer of angular momentum from a pinned neutron superfluid to the observable crust \cite{anderson,alpar,haskell,glampedakis,Antonopoulou2022}, yet the origin of the required collectivity remains debated. Avalanche models impose that collectivity through knock-on processes between vortices \cite{melatos2008,warszawski2011}, while hydrodynamic approaches obtain it from a large-scale instability of the two-fluid flow \cite{khomenko2018,antonelli2017}. The mechanisms discussed here supply it instead through the mean field itself. Collective modes provide coherent motion by construction, critical layers determine where that coherence is deposited, and the ring instability determines its azimuthal symmetry. Moreover, since the fastest collective rearrangements proceed on a fraction of a rotation period, they are comfortably compatible with the observed bounds on glitch rise times. The present model is admittedly idealized. The point-vortex gas is incompressible, two-dimensional, and contains neither explicit pinning nor mutual friction. Whether a resonant annulus actually unpins, and whether the ring instability survives the competing effects of friction and pinning, therefore lie beyond the scope of the present work \cite{graber2018,liu2025}. Yet the theory already provides a set of selection rules, namely a resonant radius fixed by the azimuthal number, a threshold ring population, a preferred multipole, and a maximum growth rate. Naturally, these ideas need not await astrophysical confirmation. Rotating quantum gases realize closely related vortex ensembles under controlled conditions, and the bridge to glitch physics has already been opened experimentally. Poli {\it et al.} observed glitch-like events in a rotating dipolar supersolid, where the density modulation pins the vortices and the release is collective \cite{poli2023}. The theory developed here describes the interval between such events, when the unpinned vortices once again form a collisionless gas. Its two sharpest predictions, namely the damping rate of the quadrupolar quasi-mode and the clump number selected by an orbiting vortex ring, are free of adjustable parameters and should be observable in the same class of quasi-two-dimensional condensates where Onsager clustering has already been reported \cite{neely,reeves,simula,bradley}. 

More broadly, the present results suggest that vortex gases belong in the same conceptual class as plasmas and self-gravitating systems. In all three cases, long-range interactions generate collective dynamics that cannot be reduced to the motion of individual constituents. From this perspective, Landau damping, phase mixing, and beam-driven instability are not peculiarities of charged particles or gravitating matter, but generic manifestations of collisionless kinetic dynamics. The point-vortex gas provides a particularly simple realization of these ideas, and offers a promising setting in which to explore them both in laboratory quantum fluids and in the superfluid interiors of neutron stars.

\appendix
\section{Dielectric response of a point-vortex gas}
\label{app:lambda_m}

In this Appendix we construct the dielectric response operator
associated with the logarithmic interaction between point vortices,
first determining the azimuthal projection of the Green function and
then obtaining the interaction kernel $\Lambda_m(J,J')$. The
angle--action decomposition employed here is standard in the kinetic
theory of inhomogeneous long-range systems
\cite{heyvaerts2010,chavanis2012aa,chavanis2024}; what differs is its
use, since we retain the response operator itself rather than the
collision kernel built from it.

The perturbed stream function satisfies
\begin{equation}
\nabla^2\delta\psi=-\Gamma\delta f,
\label{A1}
\end{equation}
whose formal solution reads
\begin{equation}
\delta\psi({\bf r})=-\frac{\Gamma}{2\pi}\int\delta f({\bf r}')\ln|{\bf r}-{\bf r}'|\,d^2r'.
\label{A2}
\end{equation}
In polar coordinates the logarithmic Green function admits the
multipolar expansion
\begin{equation}
\ln|{\bf r}-{\bf r}'|=\ln r_>-\sum_{m=1}^{\infty}\frac{1}{m}\left(\frac{r_<}{r_>}\right)^m\cos[m(\theta-\theta')],
\label{A3}
\end{equation}
with $r_<=\min(r,r')$ and $r_>=\max(r,r')$, so that, expanding the
perturbations as
\begin{equation}
\delta\psi(r,\theta)=\sum_m\psi_m(r)e^{im\theta},
\qquad
\delta f(r,\theta)=\sum_m f_m(r)e^{im\theta},
\label{A5}
\end{equation}
and projecting Eq.~\eqref{A2} onto the harmonic $m$, one obtains
\begin{equation}
\psi_m(r)=\frac{\Gamma}{2m}\int_0^\infty r' dr'\left(\frac{r_<}{r_>}\right)^m f_m(r').
\label{A6}
\end{equation}
Introducing the action variable $J=r^2/2$, for which $r'dr'=dJ'$,
Eq.~\eqref{A6} becomes
\begin{equation}
\psi_m(J)=\frac{\Gamma}{2m}\int_0^\infty\left(\frac{r_<}{r_>}\right)^m f_m(J')\,dJ',
\label{A8}
\end{equation}
and comparison with the generic integral form
\begin{equation}
\psi_m(J)=\int_0^\infty\Lambda_m(J,J')f_m(J')\,dJ'
\label{A9}
\end{equation}
identifies the interaction kernel as
\begin{equation}
\Lambda_m(J,J')=\frac{\Gamma}{2m}\left(\frac{r_<}{r_>}\right)^m =\frac{\Gamma}{2m}\left(\frac{J_<}{J_>}\right)^{m/2}.
\label{A10}
\end{equation}
The self-coupling of a resonant vortex layer follows by setting
$J=J'$, which yields
\begin{equation}
\Lambda_m(J_c,J_c) =
\frac{\Gamma}{2m},
\label{A11}
\end{equation}
showing that the interaction strength decreases with increasing
azimuthal mode number: higher-order multipolar perturbations are
therefore less effective at coupling distant vortex annuli and become
progressively more localized.

The linear response of the vortex distribution being
\begin{equation}
f_m(J)=m\frac{f_0'(J)}{\omega-m\Omega(J)}\psi_m(J),
\label{A13}
\end{equation}
substitution of Eq.~\eqref{A13} into Eq.~\eqref{A9} yields
\begin{equation}
\psi_m(J)=m\int_0^\infty\Lambda_m(J,J')\frac{f_0'(J')}{\omega-m\Omega(J')}\psi_m(J')\,dJ',
\label{A14}
\end{equation}
a Fredholm integral equation for the collective mode amplitude which
shows that the logarithmic interaction couples different values of the
action variable and therefore defines a non-local spectral problem.
Introducing the kernel of the dielectric operator,
\begin{equation}
\varepsilon_m(\omega;J,J')=
\delta(J-J')
-m\Lambda_m(J,J')
\frac{f_0'(J')}
{\omega-m\Omega(J')},
\label{A15}
\end{equation}
the mode equation may be written compactly as
\begin{equation}
\int_0^\infty \varepsilon_m(\omega;J,J')\,\psi_m(J')\,dJ' =0,
\label{A16}
\end{equation}
so that collective oscillations, being the non-trivial solutions of
this homogeneous equation, satisfy the spectral condition
\begin{equation}
\det\hat\varepsilon_m(\omega)=0.
\label{A17}
\end{equation}
%

\section{The thin-ring limit} 
\label{app:thin_ring} 

A distribution containing the ring contribution of Eq.~\eqref{eq:ring_equil} cannot be inserted directly into Eq.~\eqref{eq:Fredholm}, because it generates $\delta'(J-J_b)$ against the kernel $\Lambda_m(J,J')$, which is continuous but not differentiable at $J=J'$. We therefore regularize the ring, derive the dispersion relation at finite width, and only then take the thin-ring limit. Replace the second term of Eq.~\eqref{eq:ring_equil} by an annulus of width $\Delta$ and uniform density $n_r=N_b/\Delta$ occupying $J_1<J<J_2$, with $J_{1,2}=J_b\mp\Delta/2$. The equilibrium now contains three vorticity jumps,
\begin{equation} \frac{\partial f_0}{\partial J} = -n_0\,\delta(J-J_R) +n_r\,\delta(J-J_1) -n_r\,\delta(J-J_2), \label{eq:app_jumps} 
\end{equation} 
and the rotation frequency, obtained from the enclosed circulation, remains continuous but takes three distinct values at those jumps, 
\begin{equation} \Omega_R=\Omega_0, \qquad \Omega_1=\frac{\Omega_0J_R}{J_1}, \qquad \Omega_2=\frac{\Omega_0J_R(1+\eta)}{J_2}. \label{eq:app_omegas} 
\end{equation} 
The factor $1+\eta$ in $\Omega_2$ is the ring's own circulation, and it is the single ingredient on which everything that follows depends. Although $J_2-J_1\to0$, the two faces of the ring retain a finite frequency separation, 
\begin{equation} \Omega_2-\Omega_1 \longrightarrow \frac{\Gamma N_b}{2J_b} = \frac{\eta\,\Omega_0}{\xi^2}, \label{eq:app_sep} 
\end{equation} 
which is precisely Eq.~\eqref{eq:ring_faces} of the main text. It is convenient to introduce 
\begin{equation} 
D_0=\omega-m\Omega_R, \qquad D_-=\omega-m\Omega_1, \qquad D_+=\omega-m\Omega_2, \label{eq:app_denoms} \end{equation} 
which coincide, in the thin-ring limit, with the denominators used in Eq.~\eqref{eq:ring_cubic}. With Eq.~\eqref{eq:app_jumps}, the integral in Eq.~\eqref{eq:Fredholm} collapses onto the three values $\psi_m(J_i)$, which are well defined because $\psi_m$ remains continuous across each jump. Writing $s=(-n_0,n_r,-n_r)$ for the jump strengths and using $\Lambda_m(J,J') = \frac{\Gamma}{2m} \left(\frac{J_<}{J_>}\right)^{m/2},$ the Fredholm equation becomes the homogeneous $3\times3$ system 
\begin{equation} M_{ji} = \delta_{ji} - \frac{\Gamma}{2} \left(\frac{J_<}{J_>}\right)^{m/2} \frac{s_i}{D_i}, \qquad \mathbf M (\psi_R,\psi_1,\psi_2)^{\rm T} =0, \label{eq:app_matrix} 
\end{equation} 
whose dispersion relation is $ \det\mathbf M(\omega)=0$. The $2\times2$ block associated with the two ring faces is the only place where $\Delta^{-1}$ appears, through $n_r=N_b/\Delta$. Its determinant is 
\begin{equation} 1 +\frac{\Gamma}{2}\,n_r \frac{m(\Omega_2-\Omega_1)}{D_1D_2} -\left(\frac{\Gamma}{2}\right)^2 n_r^2 \frac{1-(J_1/J_2)^m}{D_1D_2}, \label{eq:app_ringblock} 
\end{equation} 
the second term arising from the difference between the two resonant denominators and the third from the cusp of the kernel $\Lambda_m$ between the two faces. Both contributions are singular. Equation~\eqref{eq:app_sep} gives $n_r(\Omega_2-\Omega_1)=\mathcal O(\Delta^{-1})$, while $ 1-(J_1/J_2)^m = m\Delta/J_b+\mathcal O(\Delta^2) $ implies $n_r^2\Delta=\mathcal O(\Delta^{-1})$ in the third term. Their singular parts are 
\begin{equation} 
+\left(\frac{\Gamma}{2}\right)^2 \frac{mN_b^2}{\Delta J_b D_1D_2} \qquad\text{and}\qquad -\left(\frac{\Gamma}{2}\right)^2 \frac{mN_b^2}{\Delta J_b D_1D_2}, \label{eq:app_cancel} 
\end{equation} 
and therefore cancel identically. The cancellation occurs because the frequency separation of the two faces, Eq.~\eqref{eq:app_sep}, is itself generated by the ring's self-field. Omitting the ring contribution from Eq.~\eqref{eq:ring_profile} would set $\Omega_2=\Omega_1$, eliminate the first singular term in Eq.~\eqref{eq:app_ringblock}, and leave the second divergent, so that the thin-ring limit would fail to exist. Physically, an infinitesimally thin annulus carrying finite circulation is a vortex sheet, whose Kelvin--Helmholtz response remains finite rather than diverging as the layer thickness vanishes. What survives of Eq.~\eqref{eq:app_ringblock} as $\Delta\to0$ is 
\begin{equation} 1-\frac{\omega_{b,0}^2}{D_-D_+}, \qquad \omega_{b,0}^2 = \frac{m\eta}{\xi^4} \left[ 1-\frac{(m-1)\eta}{2} \right] \Omega_0^2, \label{eq:app_bare} 
\end{equation} 
which is the {\it bare} ring response. It is the counterpart of the dressed quantity $\omega_b$ introduced in Eq.~\eqref{eq:ring_wb}, prior to the reaction of the patch. The patch contribution may be written as $1+\Omega_0/D_0$, while the remaining entries of Eq.~\eqref{eq:app_matrix} couple patch and ring through $ \Lambda_m(J_R,J_b) = \frac{\Gamma}{2m\,\xi^m}$, so that the loop patch $\rightarrow$ ring $\rightarrow$ patch enters quadratically as $\xi^{-2m}$. Expanding the determinant and collecting terms according to their frequency dependence gives 
\begin{equation} \det\mathbf M = \left(1+\frac{\Omega_0}{D_0}\right) \left(1-\frac{\omega_{b,0}^2}{D_-D_+}\right) +\frac{\delta\,\Sigma_m+\Omega_0 g_m^2} {D_0D_-D_+}, 
\label{eq:app_loop} 
\end{equation} 
where $\delta=\omega-(m-1)\Omega_0 $,  
and 
\begin{equation} \Sigma_m = \frac{m\eta}{\xi^{2m+2}}\Omega_0^2, \qquad g_m^2 = \frac{m(m-1)\eta}{\xi^{2m+4}} \left( \xi^2-1-\frac{\eta}{2} \right)\Omega_0^2. 
\label{eq:app_sigma} 
\end{equation} 
The loop contribution therefore separates into a part proportional to $\delta$ and a part independent of frequency. Using 
\[ 1+\frac{\Omega_0}{D_0} = \frac{\delta}{D_0}, \] 
the term $\delta\Sigma_m/(D_0D_-D_+)$ can be absorbed into the second factor of Eq.~\eqref{eq:app_loop}, and Eq.~\eqref{eq:ring_cubic} follows with 
\begin{equation} \omega_b^2 = \omega_{b,0}^2-\Sigma_m = \frac{m\eta}{\xi^4} \left[ 1-\xi^{-2(m-1)} -\frac{(m-1)\eta}{2} \right] \Omega_0^2, 
\label{eq:app_dressed} 
\end{equation} 
which is Eq.~\eqref{eq:ring_wb} of the main text. The quantity appearing in the final dispersion relation is therefore not the bare ring response $\omega_{b,0}$ but the dressed quantity $\omega_b$, the geometric factor $1-\xi^{-2(m-1)}$ originating from the field raised by the ring on the patch edge and subsequently scattered back onto the ring with opposite sign. The second part of the loop, represented by $g_m^2$, is the resonant coupling coefficient appearing in Eq.~\eqref{eq:ring_g}. It survives on the resonant circle: at $\xi=\xi_{\rm res}$ one has $(m-1)(\xi^2-1)=1$, so that $g_m^2 \to m\eta\Omega_0^2/(\xi^{2m+4})$ as $\eta\rightarrow 0$. Equations~\eqref{eq:app_loop}--\eqref{eq:app_dressed} have been checked against direct evaluation of $\det\mathbf M(\omega)$ from Eq.~\eqref{eq:app_matrix} at finite $\Delta$ in extended-precision arithmetic. Such precision is required because the cancellation in Eq.~\eqref{eq:app_cancel} subtracts quantities of order $\Delta^{-1}$ to leave an $\mathcal O(1)$ remainder. The agreement is better than fourteen significant figures for $m=2,3,4$ within $0.05\le\eta\le0.7$ and $1.15\le\xi\le1.6$.

\begin{acknowledgements} 

This paper is based upon work from COST Action SCALES CA24139, supported by COST (European Cooperation in Science and Technology).

\end{acknowledgements}

\bibliographystyle{apsrev4-1}
\bibliography{references.bib}

\end{document}